\documentclass[twocolappendix,numberedappendix]{emulateapj}

\newcommand{\be}{\begin{equation}}
\newcommand{\ee}{\end{equation}}

\newcommand{\Lya}{Ly$\alpha$}

\usepackage{color, soul}
\usepackage{ulem}
\usepackage{graphics}
\usepackage{subfigure}
\usepackage{amsmath}
\usepackage{xfrac}

\shorttitle{3D Simulated Hot Jupiter Mass Loss}
\shortauthors{Tripathi et al.}

\begin{document}

\title{Simulated Photoevaporative Mass Loss from Hot Jupiters in 3D}

\author{
Anjali Tripathi \altaffilmark{1}, 
Kaitlin M.~Kratter \altaffilmark{2}, 
Ruth A.~Murray-Clay \altaffilmark{1,3},
Mark R.~Krumholz \altaffilmark{4}
}

\altaffiltext{1}{Harvard-Smithsonian Center for Astrophysics, 60 Garden Street, Cambridge, MA 02138, USA;
{\tt atripathi@cfa.harvard.edu}}
\altaffiltext{2}{Department of Astronomy and Steward Observatory, University of Arizona, Tucson, AZ 85721, USA}  
\altaffiltext{3}{Department of Physics, University of California, Santa Barbara, CA 93106, USA}  
\altaffiltext{4}{Department of Astronomy and Astrophysics, University of California, 1156 High Street, Santa Cruz, CA 95064, USA}

\begin{abstract}
Ionizing stellar photons heat the upper regions of planetary atmospheres, driving atmospheric mass loss.  Gas escaping from several hot, hydrogen-rich planets has been detected using UV and X-ray transmission spectroscopy.  Because these planets are tidally locked, and thus asymmetrically irradiated, escaping gas is unlikely to be spherically symmetric.   
In this paper, we focus on the effects of asymmetric heating on local outflow structure.  We use the Athena code for hydrodynamics to produce 3D simulations of hot Jupiter mass loss that jointly model wind launching and stellar heating via photoionization. 
Our fiducial planet is an inflated, hot Jupiter with radius $R_{\rm p}=2.14 R_{\rm Jup}$ and mass $M_{\rm p} = 0.53 M_{\rm Jup}$.  We irradiate the initially neutral, atomic hydrogen atmosphere with 13.6~eV photons and compute the outflow's ionization structure.  There are clear asymmetries in the atmospheric outflow, including 
a neutral shadow on the planet's nightside.  Given an incident ionizing UV flux comparable to that of the Sun, we find a steady-state mass loss rate of $\sim$$2\times10^{10}$ g s$^{-1}$.  The total mass loss rate and the outflow substructure along the substellar ray show good agreement with earlier 1D models, for two different fluxes.
Our 3D data cube can be used to generate the outflow's extinction spectrum during transit.  As a proof of concept, we find absorption of stellar \Lya~at Doppler-shifted velocities of up to 
$\pm 50$~km~s$^{-1}$. Our work provides a starting point for further 3D models that can be used to predict observable signatures of hot Jupiter mass loss.
\end{abstract}

\keywords{hydrodynamics -- planets and satellites: atmospheres, gaseous planets -- planet-star interactions}


\section{Introduction}

Photoionization by high energy stellar radiation heats the upper layers of planetary atmospheres.  This heating drives thermal mass loss and can thus play a substantial role in a planet's atmospheric evolution.  Strongly irradiated 
hydrogen-rich planets are most susceptible to mass loss, 
and transit observations of close-in giant planets at UV and X-ray wavelengths have revealed atmospheric escape.  However, models are required to translate these observations into constraints on mass loss rates or outflow structure.  While models have been developed for hot Jupiter mass loss, none have yet consistently modeled the heating and three-dimensional (3D) structure of atmospheric escape.  The inherent asymmetry in the physics of atmospheric escape, especially asymmetric irradiation expected of tidally locked hot Jupiters, necessitates full 3D modeling.  To examine how asymmetric heating affects the structure of the outflow near the planet, we develop a 3D, self-consistent model of mass loss driven by photoionization heating.

The first indication of hot Jupiter mass loss came from the transmission spectroscopy of \citet{vidal03}, who observed decreased stellar \Lya~emission during transits of the hot Jupiter, HD 209458b.  The depth of the transit in \Lya, $15\%\pm 4\%$, was ten times larger than the optical transit depth of 1.5\% \citep{charbonneau, henry}, suggesting absorption by a neutral hydrogen atmosphere larger than the planet's Roche lobe.  Absorption out to Doppler equivalent velocities of $\pm100$~km~s$^{-1}$ indicated that the neutral gas either moved at high velocities or had a large column depth, enhancing the line's naturally broadened wings.  
Subsequent studies have confirmed stellar  \Lya~absorption by extended gas around HD 209458b \citep{ehrenreich} and the hot Jupiter HD 189733b \citep{lecavelier10}, with temporal variations in the measured absorption \citep{benjaffel, lecavelier12}.
Absorption in CII, OI,  and SiIII for HD 209458b (\citealp{vidal04, linsky}; but see \citealp{ballester}) and in X-rays for HD 189733b \citep{poppenhager} suggest that the atmospheric outflows of these planets are metal enriched, although interpretation of the X-ray signal may be complicated by stellar variability \citep{llama15}.  Metal line absorption has also been detected for the hot Jupiter WASP-12b \citep{haswell}, which is thought to be overflowing its Roche lobe \citep{li10}.
Stellar \Lya~absorption observations of a transiting hot Neptune, GJ 436b, suggest it has an asymmetric outflow structure \citep{kulow, ehrenreich15}.  

Stellar heating induces planetary mass loss through hydrodynamic escape.  This heating is primarily due to extreme UV stellar radiation, which photoionizes neutrals in the  planet's upper atmosphere, liberating electrons that heat the gas through collisions.  
This photoionization heating creates pressure gradients that accelerate the gas from subsonic to supersonic velocities.  Consequently, heated gas moves outward in the form of a planetary wind, similar to the structure of the solar wind \citep{parker}.  Gas exceeding the escape velocity or displaced beyond the Roche lobe becomes unbound and escapes into space.  
For the large ionizing fluxes received by hot Jupiters, hydrodynamic escape is the most efficient mass loss process.  Other mass loss mechanisms, including Jeans escape \citep{chamberlain}, are less efficient because they operate on individual particles, rather than a collective fluid.

One-dimensional (1D) hydrodynamic models of energy-limited escape from hot Jupiters were first calculated by \citet{lammer} and \citet{baraffe} to explain the \citet{vidal03} observations of HD 209458b.  Based on work done for the early Earth and Venus \citep{gross, watson}, these pioneering calculations assumed that all energy deposited by stellar radiation goes into heating.  The heat is conducted to lower radii and drives gas to larger radii, where it can absorb more flux, thus enhancing the outflow.  These models predicted catastrophic evaporation for hot Jupiters.  Other energy-limited models \citep[including][]{hubbard} and more detailed 1D models which have accounted for chemistry, heating and cooling, tidal gravity, and the stellar wind \citep[][hereafter MC09]{yelle, garciamunoz, murray} have suggested less dramatic mass loss for hot Jupiters.  Hot Jupiters, unlike the early Earth and Venus, do not efficiently conduct heat downward \citep[MC09]{garciamunoz}.

Not modeled by 1D studies are the inherent asymmetries in atmospheric escape processes.  The stellar wind's pressure confinement, rotation from the Coriolis force, and magnetic fields add to the asymmetry of escaping atmospheric gas.  
Day and night differences due to the stellar wind have been captured by mass loss models in 2D \citep{proga, tremblin} and in 3D \citep{bisikalo, llama}.   These and other 3D simulations \citep{schneiter, cohen, bourrier, matsakos}, which variously include orbital motion, radiation pressure, or magnetic fields, have not directly included the planetary wind's production by ionizing radiation from the host star.  
Instead, they have initialized the temperature at the wind base and used this to generate a planetary wind.  
The recent work of \citet{owenadams} moves toward a more self-consistent picture by simulating ionization-driven winds from hot Jupiters, with stellar winds and magnetic fields, in 2D.  Still, there is a need for 3D simulations which take into account photoionization, and this motivates our work. 

As a first step toward realistic models of hot Jupiter mass loss, we present a new 3D hydrodynamic model of atmospheric escape with self-consistent heating.  We focus on how asymmetric heating affects the flow near the planet.  The model and results are organized in this paper as follows.  In Section \ref{sec:model}, we describe the physics included in our simulation.  In Section \ref{sec:methods} we describe our simulation setup and our initial conditions.  In Section \ref{sec:results}, we present our results of the time-evolved wind structure, mass-loss rates, and comparisons to 1D models.  In Section \ref{sec:obs}, we provide our estimate of the predicted~\Lya~transmission spectrum mid-transit.  In Section \ref{sec:discussion}, we conclude with a summary and a discussion of future extensions.

 \section{Modeling Hydrodynamic Escape}
 \label{sec:model}
 To model hydrodynamic escape we conduct 3D radiation hydrodynamic simulations.  Although the upper atmosphere of a hot Jupiter is low density, the mean free path remains small compared to the scale height, justifying the fluid approximation.  To attain sufficient dynamic range to resolve the upper atmosphere, we include only the planet in our simulation domain. The star resides outside of the computational boundary and exerts its influence through the gravitational potential and through ionizing photons entering a single boundary.    

We use the publicly available grid-based code Athena, version 4.1 \citep{stone}, to solve the equations of ideal hydrodynamics. We implement an additional module for photoionizing radiation from \cite{krumholz07f}, as described in Appendix \ref{sec:appendix}.  Our modified version of Athena, with initial conditions files, is freely available for download and use.\footnote{\url{https://github.com/tripathi/Atmospheric-Athena} }
In this section, we describe the numerical implementation of our model, and our initial conditions can be found in Section \ref{sec:methods}.

\subsection{Hydrodynamics}
\label{sec:hydro}
We use Athena to solve the following set of hydrodynamic equations, including  gravitational, radiative, and chemical evolution source terms:
\begin{eqnarray}
{\partial \rho \over{ \partial t}} + \nabla \cdot (\rho \mathbf{v}) &=&0, \\
{\partial  \over{ \partial t}} (\rho \mathbf{v})  + \nabla \cdot (\rho \mathbf{v} \mathbf{v}) + \nabla P  &=&- \rho \nabla \Phi,\\
{\partial  E \over{ \partial t}} +  \nabla \cdot ((E + P ) \mathbf{v}) &=& \mathcal{G-L}, \label{eqn:energy}\\
{\partial \rho_n \over{ \partial t}} + \nabla \cdot (\rho_n \mathbf{v}) &=& \mathcal{R-I} \label{eqn:advection}, 
 \end{eqnarray}
where $\rho$ is the total density, $\rho_n$ is the density of neutral gas, $\mathbf{v}$ is the velocity, $P$ is the thermal pressure, and $\Phi$ is the gravitational potential.  The total energy density $E$, excluding the chemical potential energy,  is
\begin{equation}
E \equiv \epsilon + \rho {\mathbf{v} \cdot \mathbf{v} \over 2}, 
\end{equation}
where $\epsilon$ is the internal energy density excluding the chemical potential.  Omitting the chemical potential allows us to use the usual relationship between pressure and internal energy for an ideal gas, 
\begin{equation}
\epsilon = {P\over {\gamma -1}},
\end{equation}
and to adopt an adiabatic equation of state with $\gamma = 5/3$ as a constant, appropriate for either the atomic or ionized gas expected to be found in the upper atmosphere of a hot Jupiter.  Excluding the chemical energy and treating $\gamma$ as constant in this manner is reasonable because we are in a regime where collisional ionization is negligible, and thus there is a strong separation of scales between the mean particle kinetic and chemical energies.  In the energy equation, Equation \ref{eqn:energy}, $\mathcal{G}$ and $\mathcal{L}$ are the rates of radiative heating and cooling, respectively.  In the continuity equation for the neutral density, Equation \ref{eqn:advection}, $\mathcal{R}$ and $\mathcal{I}$ are the rates of recombination and ionization, respectively.

\subsection{Gravitational potential}
\label{sec:gravpot}
We employ a static gravitational potential from both the planet and the host star, including the contribution from the centrifugal or tidal term:
\be
\Phi =  - \frac{GM_{\rm p}}{r} - \frac{GM_\star}{r_{\star}} - {1\over2} {GM_{\star} r_{\star}^2\over a^3} ,
\ee
where $M_{\rm p}$ is the mass of the planet, $M_{\star}$ is the mass of the star, $a$ is the distance between the centers of the planet and star, $r$ is the local distance to the planet, and $r_{\star}$ is the  local distance to the star.

\subsection{Ionization balance}
\label{sec:ioniz}
As noted in Section \ref{sec:hydro}, we track both the neutral and the ionized gas in the computational domain. We model the changes in neutral density based on rates of photoionization and recombination.  The former depends on the available stellar ionizing flux, while the latter depends on the number densities of ions and neutrals in the gas.  

We consider a simulation box whose distance from the star is significantly larger than the size of the box, $l_{\mathrm{box}}$.  We neglect geometric spreading of the stellar radiation field and simply treat it as a planar front of radiation entering the box.  Since the near edge of the box is 4.5 $l_{\mathrm{box}}$ from the star and the far edge is 5.5 $l_{\mathrm{box}}$, the change in flux between the near and far edges of the box and the center, where we have normalized, is ~20\%.  We are therefore making an error of this order by neglecting geometric spreading.   A planar front of radiation enters the box with photon flux $F_0$. In this approximation, the flux at a distance ${\rm x}$ into the box is given by
\be
F({\rm x}) = F_0 e^{-\tau({\rm x})},
\label{eqn:flux}
\ee  
 which depends only on the optical depth $\tau$, defined as 
\be
\tau ({\rm x})= \int_{0}^{\rm x} n_{\rm H} \sigma_{\rm ph}   d\ell,
\label{eqn:optdepth}
\ee
where $\sigma_{\rm ph} $ is the photoionization cross section and $n_{\rm H}$ is the neutral number density.  

Given the abundance of hydrogen in hot Jupiter atmospheres, we consider a pure hydrogen atmosphere so that $n_{\rm H} = \rho_n / \mu_{\rm H}$, where the mean gas mass per hydrogen nucleus is $\mu_{\rm H} = m_{\rm H}=1.67\times 10^{-24}$~g and $m_{\rm H}$ is the mass of a hydrogen atom.  For hydrogen photoionization at the threshold energy of 13.6~eV,  the cross section is $\sigma_{\rm ph}=6.3\times 10^{-18}$~cm$^2$.  
In reality, the cross section will depend on the photon frequency as roughly $\nu^{-3}$; however, since the chemical state is mostly controlled by photons near the ionization threshold for optically thin gas and the energetics are controlled by photons that span much less than a factor of 2 in frequency, we simply adopt the threshold cross section for all purposes.  While Equation \ref{eqn:optdepth} is true for any frequency, we assume that our ionizing source is monochromatic, for comparison with previous work.  We review our choice of monochromatic flux in Section \ref{sec:discussion}.  The calculated flux is used to determine the 
time rate of change of the ionized density, in units of g~cm$^{-3}$~s$^{-1}$, as
\be
\mathcal{I}_{\rm ph} =  \sigma_{\rm ph} \rho_n F({\rm x}).
\ee
The photoionization rate is simply $\mathcal{I}_{\rm ph} / \mu_{\rm H}$.

Our treatment of photoionization and recombination assumes the case-B condition and the on-the-spot approximation, so that the gas is optically thick to ionizing photons in regions where we consider recombination.  For recombinations to the ground level, we assume that emitted ionizing photons will ionize at the same location, effectively canceling out that recombination.  Thus, the time rate of change of the recombined density, in units of g~cm$^{-3}$~s$^{-1}$, is given by
\be
\mathcal{R} = \mu_{\rm H}\alpha^{(B)}n_{\rm e}n_{{\rm H}^+}, 
\ee
where $\alpha^{(B)}=2.59 \times 10^{-13} (T/10^4$~{\rm K})$^{-0.7}$~cm$^{3}$~s$^{-1}$ is the case-B recombination coefficient \citep{osterbrock}, $n_{{\rm H}^+}$ is the number density of protons, and $n_{\rm e}$ is the number density of electrons.  We assume that $n_{\rm e}=n_{{\rm H}^+} = (\rho - \rho_n) / \mu_{\rm H}$.  Note that the recombination rate is simply $\mathcal{R}/\mu_{\rm H}$.

We omit collisional ionization from our treatment because it is negligible for our problem, since MC09 found that temperatures at the atmosphere's wind base are $\sim 10^4$~K.  The gas temperature $T$ is determined from the total energy density $E$ and momentum density ${\bf p}$ using
\be
T = {\gamma-1 \over \rho} \left (E -{1 \over 2} {|{\bf p}|^2 \over \rho} \right) {\mu \over k}.
\label{eqn:temp}
\ee
where the mean gas mass, $\mu = {{x\mu_{\rm i}} + (1-x) \mu_{N}}$, depends on the ionization fraction, $x = 1 - {\rho_n /\rho}$, the mean particle mass of ionized species $\mu_{\rm i}$, and  the mean particle mass of neutral gas $\mu_N$.  For our hydrogen gas, we use $\mu_{\rm i} = m_{\rm H}/2$ and $\mu_N = m_{\rm H} =1.67\times10^{-24}$~g. We note that though  the true value of $\mu$ is $\mu = m_{\rm H}/(1+x)$ for atomic and ionized hydrogen, the above expression, chosen for compatibility with previous code development, provides a good approximation.

 \subsection{Heating and cooling}
 \label{sec:heating}
Radiation not only ionizes the gas, but it also contributes to its heating and cooling.  Each ionizing photon imparts its energy in excess of the ionization threshold to the newly liberated electron, which then heats the gas through collisions.  Each photon contributes an energy, $e_{\Gamma}$, to heating the gas, so that the photoionization heating rate, per unit volume, is
\be
\mathcal{G}_{\rm ph} = e_{\Gamma}\sigma_{\rm ph}n_{\rm H} F({\rm x}).
\ee
We use an ISM appropriate heating rate of $e_{\Gamma}=2.4$~eV \citep{whalen}, chosen to validate our ionizing radiative transfer code against HII region ionization fronts, discussed in Appendix \ref{sec:accuracy}.  While we use this heating rate for both our validation tests and our mass loss model, we note that for a hot Jupiter atmosphere absorbing a quiet solar Lyman continuum spectrum, \citet{trammell} calculated  $e_{\Gamma}=2.7$~eV, assuming 100\% energy deposition efficiency.  Efficiency values may be smaller \citep{koskinen}, making our choice of heating rate an upper limit for mass loss. 

Hydrogen recombination emits photons, which can escape and thus cool the gas.  The radiative recombination cooling rate, from a linear fit to values in \citet{osterbrock}, is
\be
\mathcal{L}_{\rm rec}\approx  \left(6.11 \times 10^{-10} {\rm ~ cm^{3}~s^{-1}}\right) kT \left( {T \over {\rm K}}\right)^{-0.89} n_{\rm e} n_{{\rm H}^+},
\ee
where $k$ is the Boltzmann constant. 
 
We also include the rate for cooling from neutral atoms that are collisionally excited and emit \Lya~photons, given by \citet{black} as
\be
\mathcal{L}_{\rm Ly\alpha} = \left(7.5\times 10^{-19} {\rm\,erg} {\rm\,cm}^ {-3} {\rm\,s}^{-1}\right) e^{-118348{\rm\,K}/T} n_{e}n_{\rm H} .
\ee
\citet{menager} suggest that this formula may overestimate the \Lya~cooling rate by up to an order of magnitude for hot Jupiter atmospheres, due to non-LTE and other radiative transfer effects.
As we discuss in Section \ref{sec:night}, \Lya~cooling is negligible almost everywhere for our planet's parameters, so such a difference does not change our results, but it may affect hotter and denser planets.

\subsection{Numerical Algorithm}
\label{sec:codesetup}
We use Athena's Roe's linearized Riemann solver, with default second-order spatial reconstruction of the fluid variables and the directionally unsplit corner transport upwind (CTU) integrator in 3D  \citep{stone}.  To avoid the carbuncle instability \citep{quirk}, which we find when photoionization heated gas advects around the planet and converges on the nightside, we use Athena's built-in H-correction \citep{stone}, to add dissipation when strong shocks are aligned with the grid.
Our fiducial simulation length is roughly four orbital periods, by which point the wind has reached a steady state.

\section{Initial Conditions}
\label{sec:methods} 

\begin{figure}
\plotone{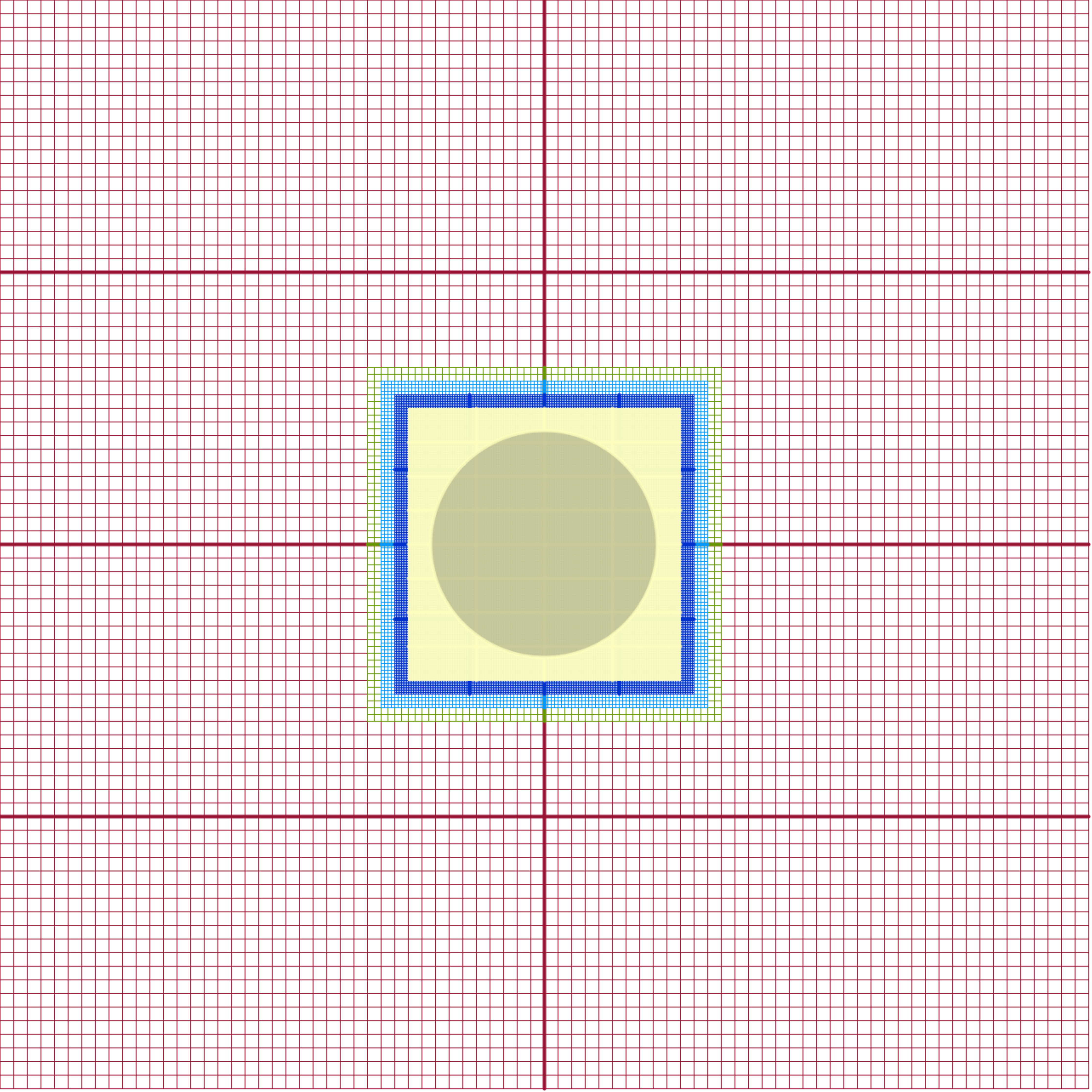}
\caption{Our computational grid has five levels of resolution (each level is in a different color), to resolve the scale height of the planet (gray).  The grid spans $1.5 \times 10^{11}$cm ($10R_{\rm p}$) in each dimension, with a highest resolution of $1/128~R_{\rm p}$.  This 2D slice in the $x-y$ plane shows how the grid is distributed across processors, as denoted by the thick lines. \label{fig:grid}}
\end{figure}

In this section, we describe our choice of initial conditions for the computational domain, planetary atmosphere, and the host star.  As shown in Figure \ref{fig:grid}, we initialize our computational domain with a planet at the origin of a 3D Cartesian grid.  The computational volume corresponds to a physical size of $(10R_{\rm p})^3$, discretized into a base grid of 80$^3$ cells.  To resolve the minimum atmosphere scale height when the wind is launched, we use five levels of grid refinement around the planet, yielding a finest resolution of $({1}/{128}~ R_{\rm p})^3$.  We show in Appendix \ref{sec:converge} that our resolution is sufficient for our results to reach numerical convergence.  At this resolution, the influence of the Cartesian discretization on the spherical atmosphere is minimal. 

We work in the frame corotating with the planet's orbit.  We assume that the planet is at a fixed distance from the star.  In this work, we omit the Coriolis force.  We do not include the rotation of the planet since hot Jupiters are tidally locked.  We inject stellar UV flux into one face ($-x$) of our box and calculate the optical depth to photoionization from this boundary.  For the fluid variables, all faces of our box have outflow boundary conditions.

\subsection{Planetary atmosphere}
\label{sec:gassetup}
To study hydrodynamic escape, we retain a large, well-resolved atmosphere to serve as a mass reservoir from which a wind can be launched. 
The atmosphere is in hydrostatic equilibrium, with a gas density profile:
\be
{dP\over dr} = -{GM_{\rm p}\rho \over r^2},
\ee
and in agreement with our adiabatic equation of state, we assume a polytrope 
\be
P = K \rho^\gamma,
\ee
where $~K$ is a constant of proportionality and $\gamma$ is the adiabatic index.  Integrating from a radius $R_{\rm p}$ (with corresponding density $\rho_{\rm p}$) to a radius $r$ yields
\be
\rho_{\rm{atm}}(r) = \left[{\gamma-1 \over \gamma}{GM_{\rm p} \over K} \left( {1\over r} - {1\over R_{\rm p}} \right) + \rho_{\rm p}^{\gamma-1}  \right]^{1/(\gamma-1)}.
\label{eqn:rhoprof}
\ee
For an ideal gas, $~K$ can be determined from the isothermal sound speed $c_{s}$ and $\rho$ using
\be
K = \rho^{1-\gamma}c_s^2.
\ee
We normalize these parameters at $R_{\rm p}$, the radius where ionizing photons are absorbed and from which the wind is launched.

\begin{deluxetable}{lc}
\tablewidth{0pt}
\tablecaption{Simulation Parameters\label{tab:params}}
\tablehead{\colhead{Parameter} & \colhead{Value}}
\startdata
\bf{Planet}& \\
Mass, $M_{\rm p}$ [g] & $10^{30}$ \\
UV photon absorption radius, $R_{\rm p}$ [cm] & $1.5\times10^{10}$ \\
Density at $R_{\rm p}$, $\rho_{\rm p}$ [g~cm$^{-3}$] & $10^{-15}$\\
Isothermal sound speed at $R_{\rm p}$, $c_{s,p}$ [cm s$^{-1}$] & $3\times10^5$\\
Hydrogen photoionization cross section, $\sigma_{\rm{ph}}$ [cm$^2$] & $6.3\times10^{-18}$\\
Mean mass per hydrogen nucleus, $\mu_{\rm H}$ [g] & $1.67\times 10^{-24}$\\
\bf{Star}& \\
Mass $M_{\star}$ [g] & $2.0\times 10^{33}$\\
Orbital distance, $a$ [cm] & $7.48\times10^{11}$ 
\enddata
\end{deluxetable}

Resolving the entire planet is computationally prohibitive and unnecessary, so we create an artificial inner boundary for the atmosphere.  We set this inner boundary at $r_{\rm b} = 0.75R_{\rm p}$, $\approx 4$ scale heights below the wind launching point, which is more than sufficient to maintain a reservoir of atmospheric gas.  Interior to $r_{\rm b}$, the density and temperature profiles are fixed at their hydrostatic values at every timestep, but non-zero fluxes are permitted across the boundary.  To prevent the density from diverging at the origin, as expected from Equation~\ref{eqn:rhoprof}, we set $\rho(r\le r_0) = \rho_{\rm atm} (r_0)$ at a radius $r_0=0.5 R_{\rm p}$.  Our model is insensitive to the choice of $r_0$, so long as it is smaller than $r_{\rm b}$ by the number of cells used for the Riemann solver's reconstruction method.

To maintain the stability of the simulated atmosphere, we match the pressure at the atmosphere's outer edge to that of a stationary and uniform ambient medium.  Unlike the pressure, the density of the ambient medium is discontinuous from the planet.  To ensure that the background gas does not influence the development of winds from the planet, we require it to be at a low enough density (or equivalently, a high enough temperature) so that the inevitable accretion of material onto the planet has a minimal effect on the atmospheric structure and wind launching.  Combining these constraints, we use the following profile to specify the density of the gas in the entire domain,
\be
\begin{aligned}
\rho (r) = 
\begin{cases}
\rho_{\rm atm}(r_0)  & \qquad  r<r_0,\\
\rho_{\rm atm}(r) & \qquad r_0 \le r\le r_{e},\\
\rho_{\rm atm} (r_{\rm e}) \cdot 10^{-4} & \qquad  r>r_{\rm e},
\end{cases}\\
\end{aligned}
\ee
where $r_{\rm e}$ is the edge of the planet.  The pressure profile is  
\be
\begin{aligned}
P (r) = 
\begin{cases}
K \rho_{\rm atm}(r_0) ^\gamma & \qquad  r<r_0,\\
K \rho_{\rm atm}(r)^\gamma& \qquad r_0 \le r\le r_{\rm e},\\
K \rho_{\rm atm}(r_{\rm e})^\gamma & \qquad  r>r_{\rm e}.
\end{cases}\\
\end{aligned}
\ee
Since we model a spherical planet on a Cartesian grid, the pressure is not perfectly matched between the ambient gas and the edge of the planet's atmosphere.  Nevertheless, we find that the changes in our atmosphere over a sound crossing time are substantially smaller than the outflows that develop.  

Our fiducial parameters are summarized in Table \ref{tab:params}.  We model a low mass and extended hot Jupiter with $R_{\rm p}=1.5\times10^{10}$~cm~$=2.14R_{\rm{Jup}}$ and $M_{\rm p}=10^{30}$~g~$=0.53M_{\rm{Jup}}$.  These values are similar to WASP-17b,  the most inflated exoplanet to-date, with $R_{\rm p} = 1.97 \pm 0.06 R_{\rm{Jup}}$ \citep{bento} and $M_{\rm p} = 0.477 \pm 0.033 M_{\rm{Jup}}$ \citep{southworth}.  It is important to note that the radius that we quote as $R_{\rm p}$ for our model corresponds to the absorption radius for ionizing photons, which is larger than the observed optical planetary radius.  Our choice of a low mass, low density planet reduces the computational cost to resolve the upper atmosphere and launch an outflow.

Following the parameter study of MC09, which found that mass loss rates are insensitive to wind-base temperatures $\lesssim 10^3$~K, we set $c_{s,p}= 3\times 10^5$~cm~s$^{-1}$, or equivalently $T_p= 1.1\times10^3$~K. Our definition of $R_{\rm p}$ requires that  over a scale height evaluated at $R_{\rm p}$, $H = c_s^2/(GM_{\rm p}/R_{\rm p}^2)$, the optical depth reaches unity.   This corresponds to a number density, $n_p = 6 \times 10^8$~cm$^{-3}$.  To ensure that the gas is optically thick, we extend the planet's atmosphere  to a radius $r_{\rm e}=1.02R_{\rm p}$, where $\rho(r_{\rm e})=\rho_{\rm p} /10$.
We begin with a neutral hydrogen atmosphere with the density at $R_{\rm p}$,  $\rho_{\rm p} = \mu_{\rm H}n_p=10^{-15}$~g~cm$^{-3}$.  We note that though we set our initial conditions so that initially $\tau \approx 1$ at $R_{\rm p}$, 
 the gas is allowed to self-consistently choose where the $\tau = 1$ surface lies.  While the planet begins completely neutral, the background gas is fully ionized. Starting with a fully ionized and optically thin background allows us to track the evolution of the planet alone.

\subsection{Stellar radiation}\label{sec:stellar}
We include the host star's gravity in our static gravitational potential.  We set the mass and radius of the star as $M_{\star}=M_{\odot}$ and $R_{\star} = R_{\odot}$.  The stellar radius is used to calculate the predicted \Lya~extinction during transit.  Typical of hot Jupiters, the star is located at a distance of 0.05~AU from the planet.

Since hot Jupiters are tidally locked to their host stars, stellar flux is constantly received by the same face of the planet.  We include the stellar flux as a plane-parallel source of ionizing radiation incident on one side of our computational box.  As described in Section \ref{sec:ioniz}, the plane-parallel approximation is justified by the relative sizes and orbital separation of the planet and the star.  We treat the stellar flux as a monochromatic source, rather than a full stellar spectrum.  This choice allows us to make direct comparisons with the MC09 1D model, which also uses a monochromatic source.  

We illuminate the planet gradually, slowly increasing the flux over two orders of magnitude, using the function,
\be
{F(t) \over F_0} =  5~  {\rm erf} \left ({t \over 8 \times10^4{~\rm s}}  -1.5 \right )  +5.1,
\ee
for a given choice of $F_0$.  The ``ramp up" timescale is chosen to be significantly longer than the advection time for gas moving around the planet.  By increasing the photon flux slowly, we allow the system to gradually relax to equilibrium.  
 
To examine changes in mass loss as a function of flux, we run our simulation with two different stellar flux values.  For our fiducial model, we use a maximum steady state photon flux of $2.02\times10^{14}$~cm$^{-2}$~s$^{-1}$, which corresponds to 10.1$F_0$ for $F_0=2\times10^{13}$~cm$^{-2}$~s$^{-1}$.  Since this value of $F_0$ is comparable to the solar UV flux from \citet{woods} scaled to our orbital distance of 0.05~AU, our fiducial model represents a relatively younger, hotter star than the Sun.  To study a Sun-like star, we also run a second model, where $F_0=2\times10^{12}$~cm$^{-2}$~s$^{-1}$.  Because we use single frequency photons, there is not a one-to-one correspondence between our choice of $F_0$ and the solar flux; a different frequency range would necessitate a different $F_0$.

\section{Results}
 \label{sec:results}
The structure of the planetary outflow is described in Section \ref{sec:struc}, with winds on the day and night sides described in \ref{sec:day} and \ref{sec:night}, respectively.  The mass loss rate for our fiducial model, as well as the lower-flux model, is presented in \ref{sec:massloss}.   Agreement between our simulation and 1D models is discussed in \ref{sec:1d}.

\subsection{Wind structure}\label{sec:struc}
\begin{figure*}
\begin{centering}
\includegraphics[width=\linewidth]{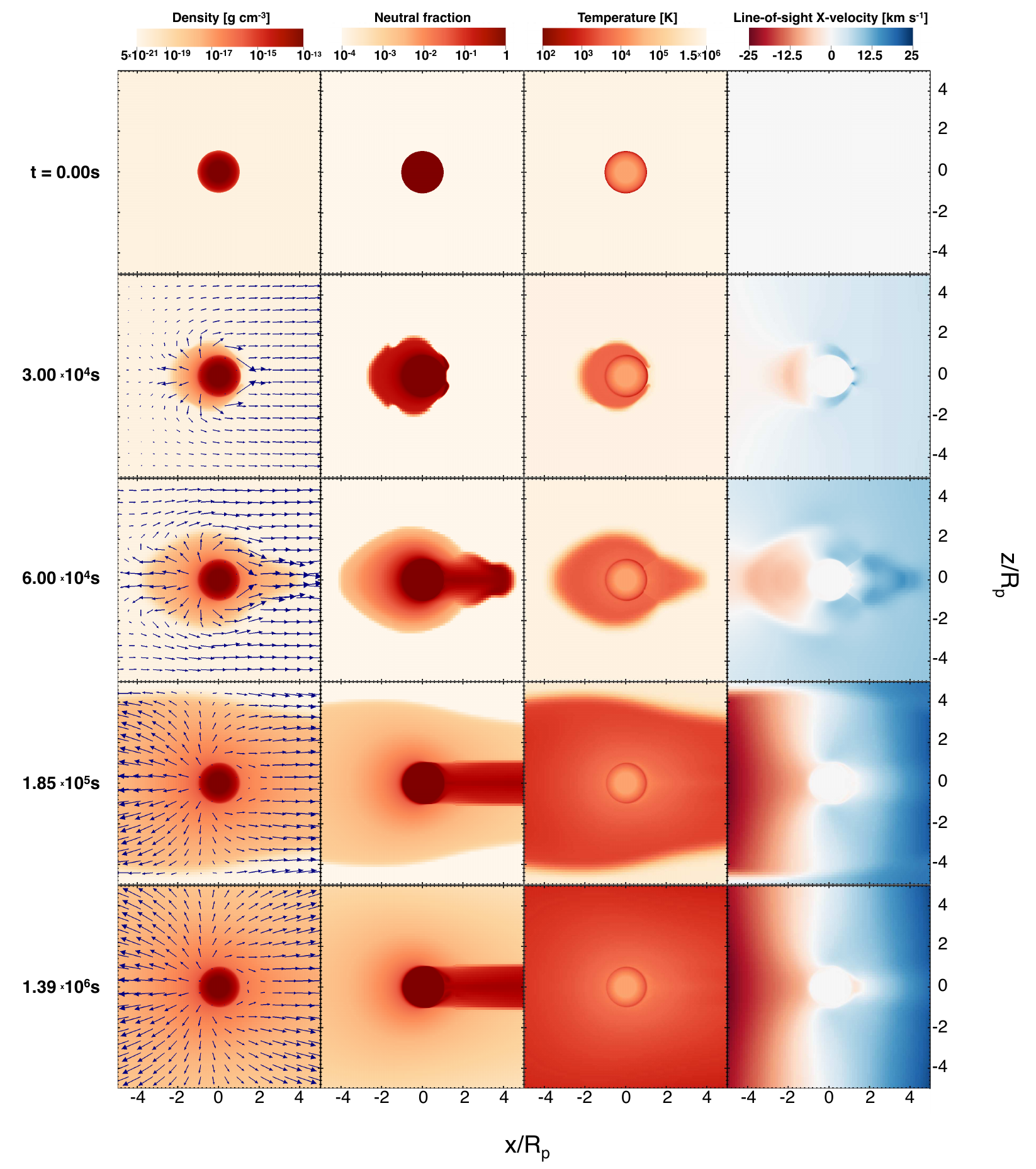}
\caption{Atmospheric expansion on the dayside, with advection toward the nightside, as seen in this time evolution of density [g~cm$^{-3}$] with velocity vectors overplotted (left), neutral fraction (second column), temperature [K] (third), and $x$-velocity [km~s$^{-1}$]  (right), for a slice through the midplane.  Top row (t=0s): initial conditions, second row ($3\times10^4$s): the atmosphere heats on the dayside and advects to the nightside, third row ($6\times10^4$s): the outflows are tidally extended along the axis to the star, fourth row ($1.85\times10^5$s): the outflows continue at larger radii, with lower density, and finally bottom row ($1.39\times10^6$s): the steady state outflow.  The star illuminates the planet from the $-x$ boundary, and the mid-transit observer is at $+x$.\label{fig:wind}}
\end{centering}
\end{figure*}

Within an orbital period of illumination, the planet develops a steady-state, transonic atmospheric outflow.  After reaching steady state at $\sim 3 \times 10^5$s, escape continues in steady state for the remainder of our simulation: three orbital periods or four sound-crossing times for our simulation box.  The transition to steady-state can be seen in the time evolution of the mass loss rate, discussed in Section \ref{sec:massloss}.  

The wind's evolution from initialization to steady state is illustrated in Figure \ref{fig:wind}.  This and other midplane visualizations were generated using VisIt \citep{visitref}.  Since our model is axisymmetric, the outflow structure is the same in both the $x-z$ (shown here) and the $x-y$ planes.  The dayside exhibits a strong, ionized outflow.  The nightside outflow is slightly weaker, with neutral gas in the planet's shadow.
We discuss these day-night differences and the underlying energetic and ionization processes in Sections \ref{sec:day} and \ref{sec:night}.  

\begin{figure}
\includegraphics[width=\linewidth]{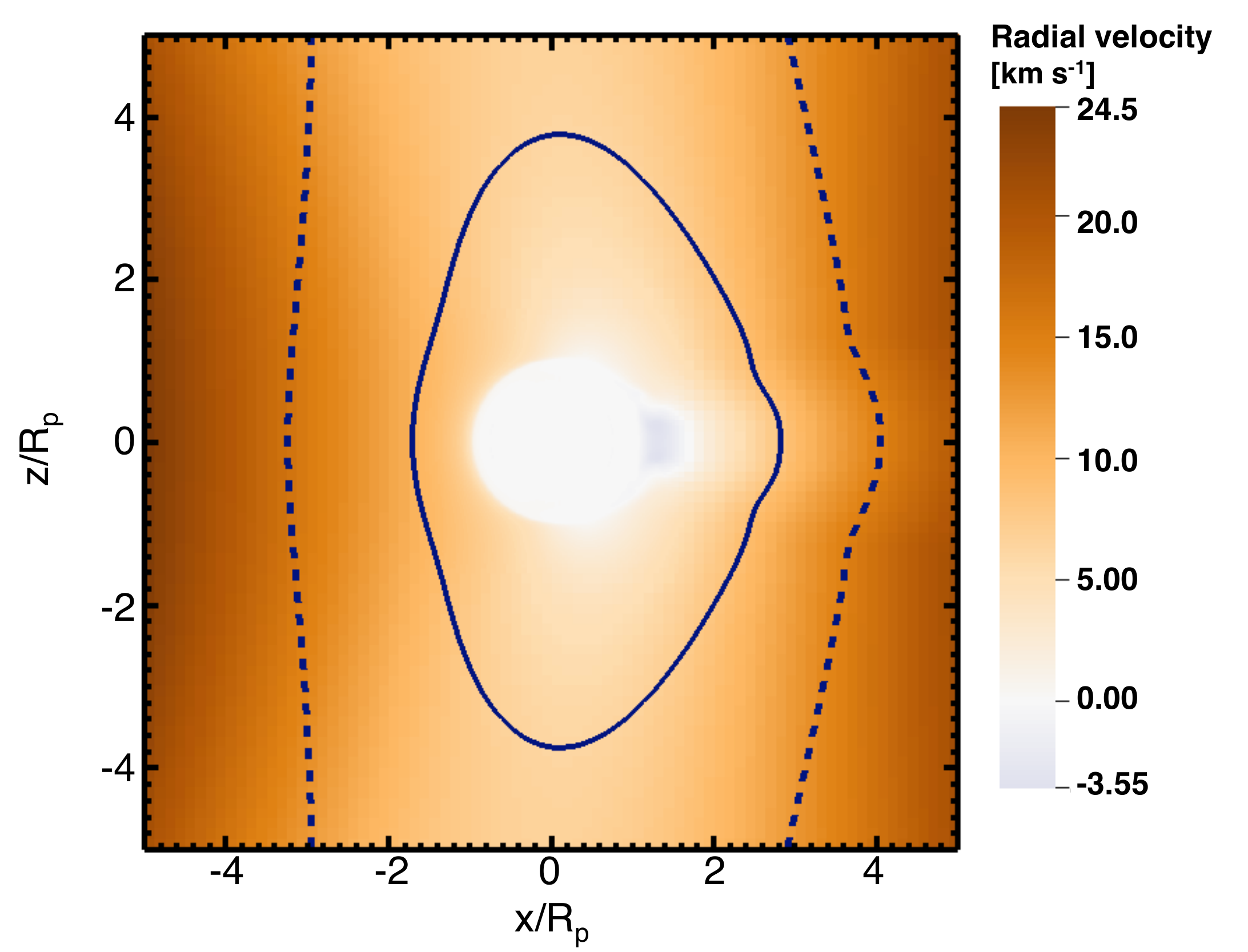}
\caption{Steady-state outflow radial velocity [km~s$^{-1}$], shown with the adiabatic sonic surface (solid) and the escape surface (dashed) in the midplane.  For reference, the Roche lobe radius is $2.7R_{\rm p}$.  \label{fig:sonic}}
\end{figure}

As shown in Figure \ref{fig:sonic}, the steady state outflow is transonic.  Its radial velocity reaches the local escape speed after exceeding the adiabatic sound speed.  The flow becomes supersonic at distances from the planet comparable to the Roche lobe radius, $R_{\rm Roche} = [M_{\rm p}/(3M_{\star})]^{1/3}a = 4.1\times10^{10}\mathrm{cm}~=2.7R_{p}$.  The escape surface is largely outside of the Roche lobe radius and is closest to the planet along the substellar ray.
 
 Tidal gravity is responsible for elongating the outflow and contributes to the asymmetric escape surface.  Consequently, 
the velocity field of the outflow observed during a planetary transit should be dominated by the line-of-sight velocity.

Throughout the results section, we discuss the wind launching in time, so as to understand its final structure and dynamics.  Because our initial conditions do not reflect the properties of a newly formed hot Jupiter, this time evolution does not represent the early evolution of a hot Jupiter's outflow.  However, the model's time evolution may be relevant for planets orbiting very active, flaring stars, whose flare timescales can be comparable to our simulated flux ramp up time.

\begin{figure}
\centering
\includegraphics[width=.93\linewidth]{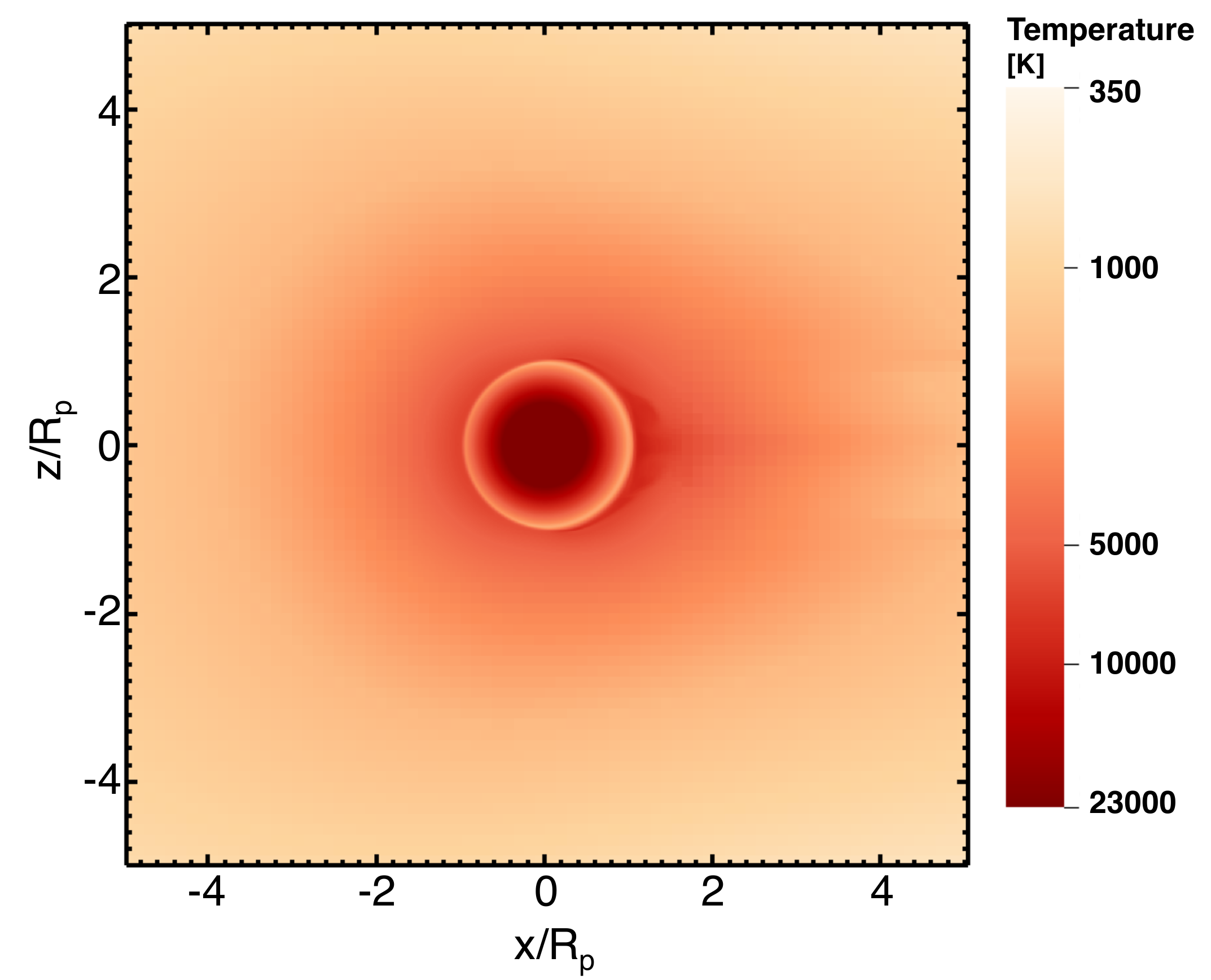}
\caption{
The temperature [K] of the steady-state outflow exhibits a day-night asymmetry, seen in this midplane slice.  While dayside temperatures remain $<10^4$~K, neutral nightside gas approaches $10^4$K.
 \label{fig:steadytemp}}
\end{figure}

\begin{figure*}
\centering
\includegraphics[width=1\linewidth]{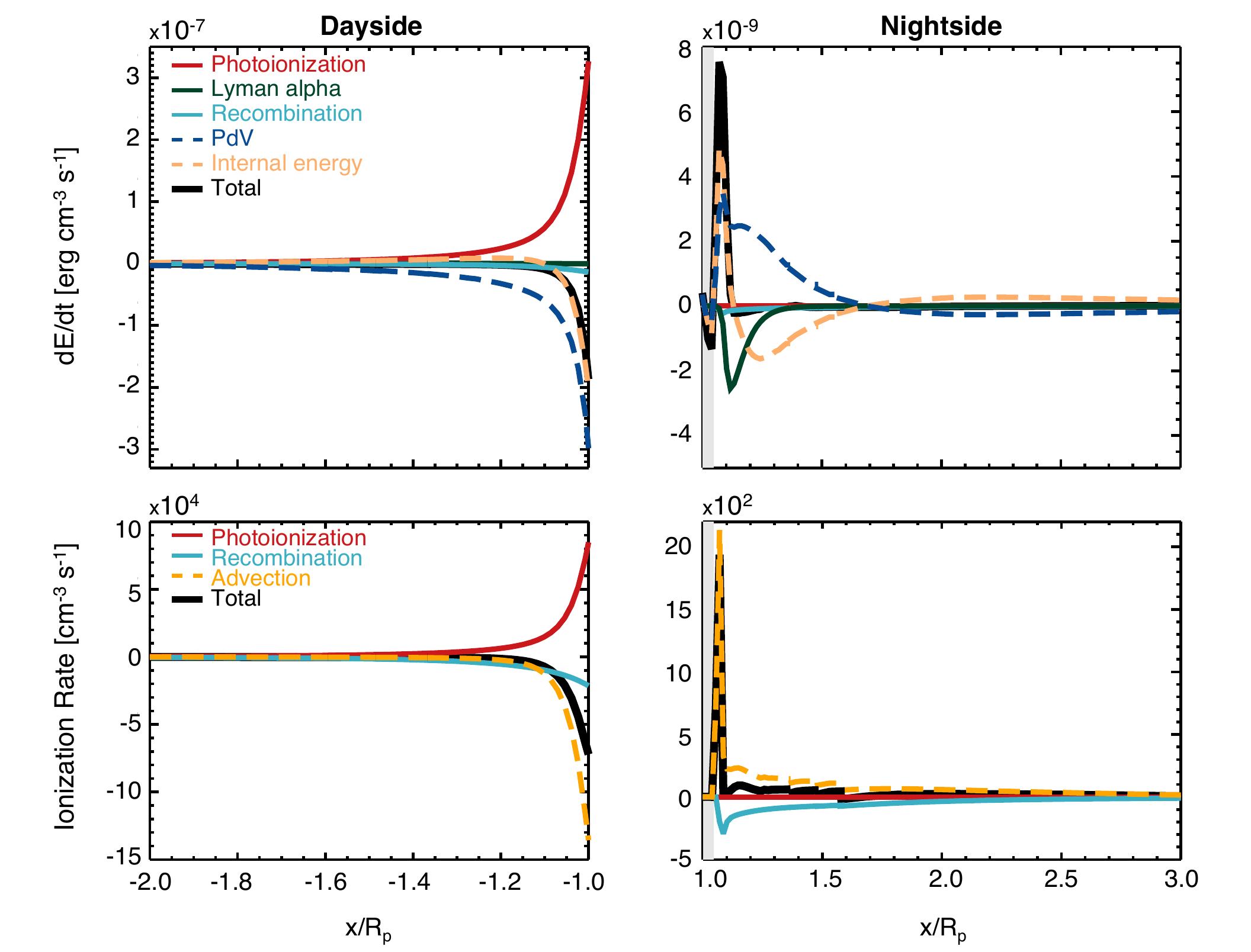}
\caption{Steady-state energy (upper) and ionization (lower) balance along the substellar ray on the dayside (left) and nightside (right), showing the terms in Equations \ref{eqn:ebalance} and \ref{eqn:ibalance}.  Above the wind base on both sides, the escaping gas is in equilibrium, as shown by the total at 0.  The shaded region highlights gas below the outflow base, on the nightside.\label{fig:balance}}
\end{figure*}

\subsubsection{Dayside flow: directly launched by photoionization heating}\label{sec:day}
On the planet's dayside, photoionization heats the atmosphere to $\sim7000$K, by depositing energy at $R_{\rm p}$, where $\tau=1$ to photoionization.  Irradiated gas above $R_{\rm p}$ accelerates and moves outwards. 
This outward expansion, hereafter referred to as PdV work, is the primary coolant of the gas. Radiative recombination cooling is an order of magnitude smaller. Because our planet has low surface gravity, 
its outflow does not achieve the $10^4$K temperature required for substantial \Lya~cooling, as shown in Figure \ref{fig:steadytemp}.  Lower temperatures are sufficient to accelerate gas to the planet's escape speed.

To assess in more detail the relative contributions of various heating, cooling, and ionization processes to the wind's structure, we recast Equations \ref{eqn:energy} and \ref{eqn:advection} as
\be \label{eqn:ebalance}
- \rho \mathbf{v} \cdot \nabla \left[ P \over \rho (\gamma -1)\right]  + {P \over \rho} \left (\mathbf{v} \cdot \nabla{\rho} \right) +\mathcal{G}_{\rm ph} - \mathcal{L}_{\rm rec} - \mathcal{L}_{\rm Ly\alpha} = 0
\ee 
and
\be  \label{eqn:ibalance}
-{\rho \over \mu}  \mathbf{v} \cdot \nabla x  +  { \mathcal{I} \over \mu_{\rm H}} -  {\mathcal{R} \over \mu_{\rm H}} = 0. 
\ee
The first two terms in Equation \ref{eqn:ebalance} represent the change in internal energy and the PdV work, respectively.  The first term in Equation \ref{eqn:ibalance} represents the advection of ions out of a given cell.

Above the wind base, PdV cooling and photoionization heating contribute to energy balance, as shown in Figure \ref{fig:balance}, which displays the terms of Equation \ref{eqn:ebalance} and \ref{eqn:ibalance} in steady state.  Near the wind base, heat is stored in internal energy, as shown by the negative
 internal energy below $1.1R_{\rm p}$ along the substellar ray.  Farther from the wind base, the outflow is driven by local photoionization heating, as well as the heat that was deposited by photoionization lower in the atmosphere and stored in internal energy.  
  
Due to the large photoionization rate of $\sim10^4$~cm$^{-3}$~s$^{-1}$, the dayside outflow is everywhere ionized. 
The dayside is in ionization balance above the wind base, as seen in the lower panel of Figure \ref{fig:balance}.  Near the wind-base, photoionization contributes a steady source of ions, which are advected away.  Ion advection is only important near the wind base, where the fraction of the gas that is ionized is low enough that recombination 
is too slow to balance ionization.  Above 1.1$R_{\rm p}$,  recombination balances photoionization.

\begin{figure}
\includegraphics[width=1\linewidth]{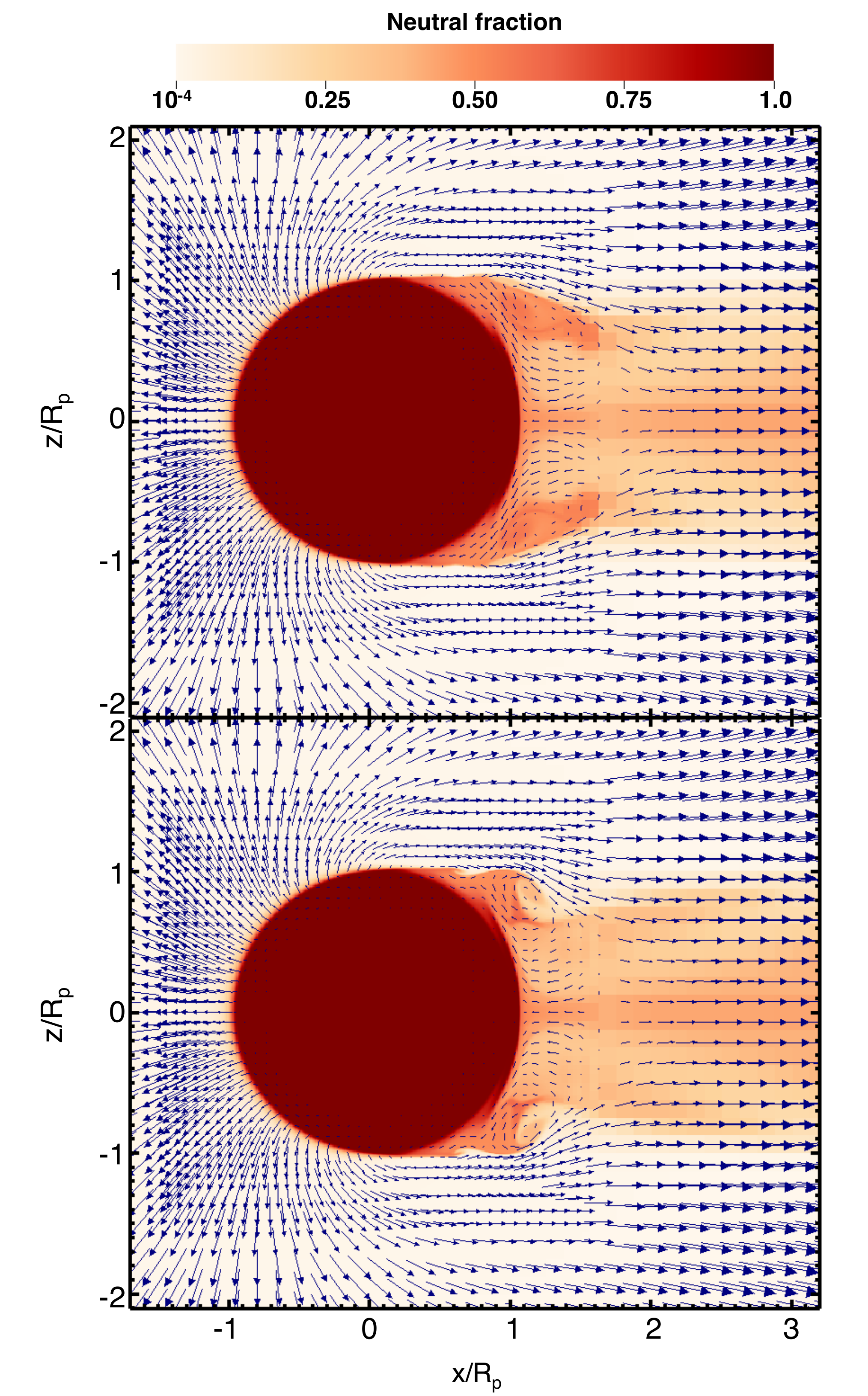}
\caption{Velocity vectors overplotted on the neutral fraction midplane slice show gas circulating on the nightside of the planet, with ever changing patterns of rolls, shown at $5\times10^5$s (upper) and $6\times10^5$s (lower).  This motion ultimately contributes neutrals to the outflow in the planet's shadow.  A stagnation point is visible along the anti-stellar ray at $\sim 1.7R_{\rm p}$, between the inward circulating material and the outflowing material.\label{fig:nightside}}
\end{figure}

\subsubsection{Nightside flow: advected ions mixed with a neutral shadow} \label{sec:night}
As seen in Figure \ref{fig:wind}, heated dayside gas not only moves radially outwards, but also advects around the planet.  The advected flow exceeds the sound speed and escape velocity, before leaving the box in a steady-state wind.  The nightside flows are aligned along the anti-stellar ray, in part due to tidal gravity.   

Unlike the dayside flow, the nightside wind is not directly driven by photoionization heating.  Instead, flows moving around the planet converge on the nightside, at a stagnation point, and heat the surrounding gas to $\sim10^4$K.  Below the stagnation point, PdV work adds to the internal energy.  Above the stagnation point, internal energy drives the outflow, which then cools by PdV expansion.  The interplay between PdV work and internal energy storage allows the gas to be in an energetic equilibrium.  With the exception of localized \Lya~cooling near the stagnation point, where the gas is hot enough, other sources of cooling, including recombination, are negligible.

In the planet's shadow, the gas has a sharply defined neutral outflow.  As highlighted in Figure \ref{fig:nightside}, the planet's nightside atmosphere contributes neutral gas to this flow, in a time-varying circulation pattern.  This gas originates above the atmosphere's initial outer radius of $1.56 \times 10^{10}$cm$ =1.04R_{\rm p}$,  rather than the dayside value of $R_{\rm p}$, since there is no photoionization in the shadow to move away material below this radius.  We note that this is the only portion of the flow that retains memory of our atmospheric initial conditions.  As shown in Figure \ref{fig:steadytemp}, this nightside neutral gas is also some of the hottest in our computational volume, due to enhanced heating from the converging flows.  

The circulation is mostly confined within a stagnation point where the advected, dayside gas converges.  Periodically the circulation cell, which may be unresolved turbulence, grows large enough so that the circulating neutral gas mixes with the advected ionized gas and then gets dragged outwards, beyond the stagnation point.  Since the planet is optically thick to photoionization, the only ionized gas in this outflow is from advection.  Outside of the planet's shadow, the gas is optically thin to photoionization and, thus, it is ionized.  Recombination increases the nightside neutral fraction, but only slightly, because the recombination timescale is long compared to the time it takes gas to flow out of our box.

\subsection{Mass loss rates and low-flux results}\label{sec:massloss}
We calculate the instantaneous mass loss rate by taking the average of the mass flux through a spherical shell, with a finite thickness of one cell, using
\be
\left< \dot{M} (r)\right> = \left< \rho {\rm v}_r \right> 4 \pi r^2.
\ee
For comparison, we also calculate the mass loss flux through the faces of a cube with dimensions $2r$ and obtain comparable mass loss rates.

\begin{figure}
\includegraphics[trim= 1cm 0cm 0cm 1cm, clip=true, width=1\linewidth]{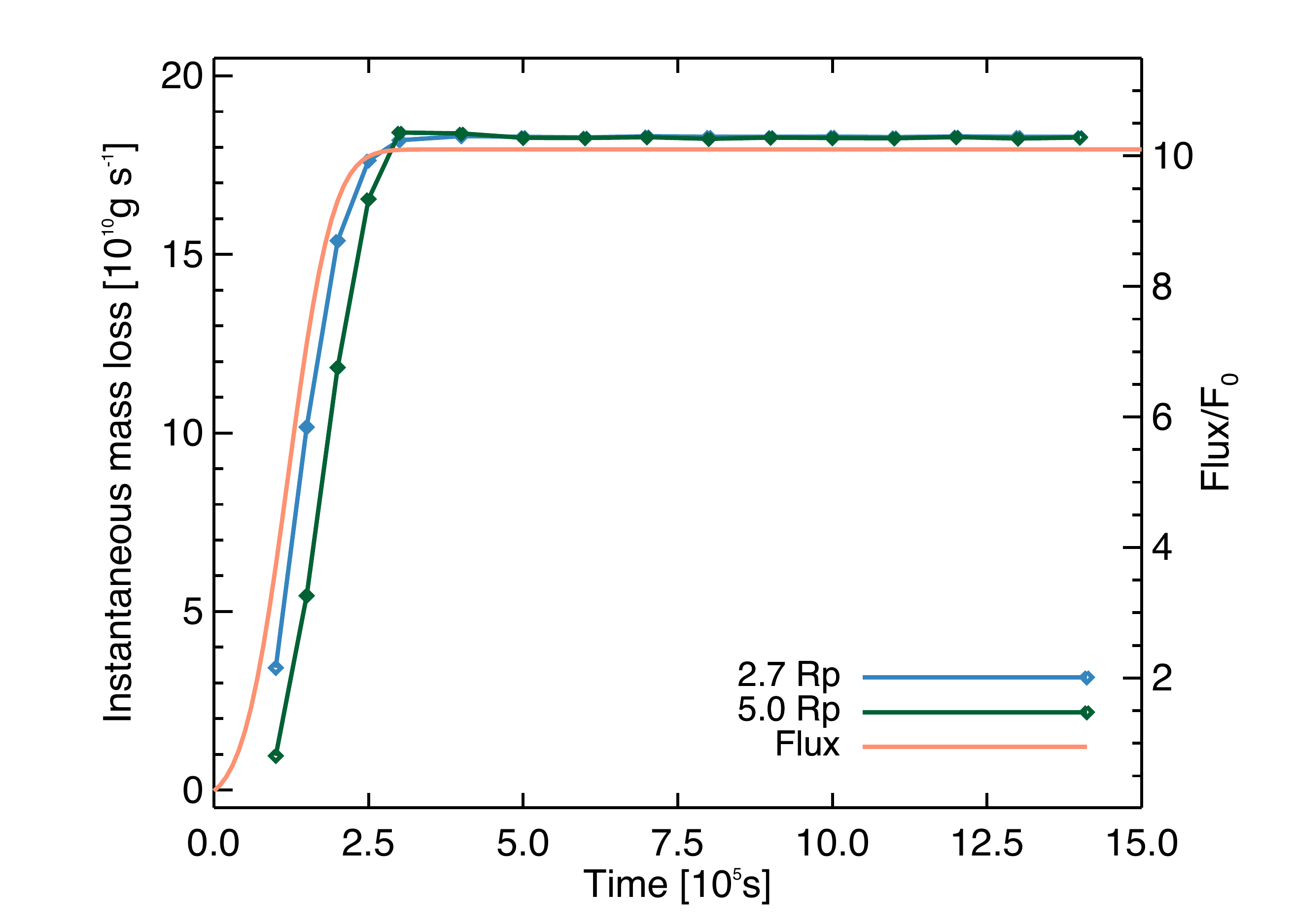}
\caption{Mass loss rates, calculated at the Roche lobe ($2.7 R_{\rm p}$, blue) and the edge of the box ($5R_{\rm p}$, green), reach a steady-state value of $1.9\times10^{11}$~g~s$^{-1}$, for a model with an incident flux of $2.02\times10^{14}$~cm$^{-2}$~s$^{-1}$.
The evolution of the mass loss rate tracks the incident UV flux (orange).  \label{fig:massloss}}
\end{figure}

As shown in Figure \ref{fig:massloss}, we find a steady-state mass loss rate of $1.9\times10^{11}$~g~s$^{-1}$ for our fiducial model.  It is interesting to note how the mass loss rate tracks the input flux over time.  Before the flux plateaus to its constant value, the mass loss rate is larger at smaller radii - indicative of the time it takes for the heated gas to expand and escape.  The mass loss rate reaches steady-state at $\sim3\times10^5$s, just after the flux reaches its constant value of $2.02\times10^{14}$~cm$^{-2}$~s$^{-1}$.  In steady-state, the mass loss rate is the same at all radii above the Roche lobe radius of $2.7R_{\rm p}$.  

To examine the mass loss rate's scaling with UV flux, we also carry out an identical simulation with lower incident flux. For a simulation with a maximum photon flux of $2.02\times10^{13}$~cm$^{-2}$~s$^{-1}$, an order of magnitude lower than our fiducial model, we find the outflow evolves with a similar structure to that of the high flux model, described in Section \ref{sec:struc}.  This lower flux simulation reaches a steady state mass loss rate of $2.2 \times 10^{10}$~g~s$^{-1}$.

\subsection{Agreement with 1D models}\label{sec:1d}
\begin{figure}[h]
\includegraphics[trim= 1.cm 1cm 1.cm 1.cm, width=1\linewidth]{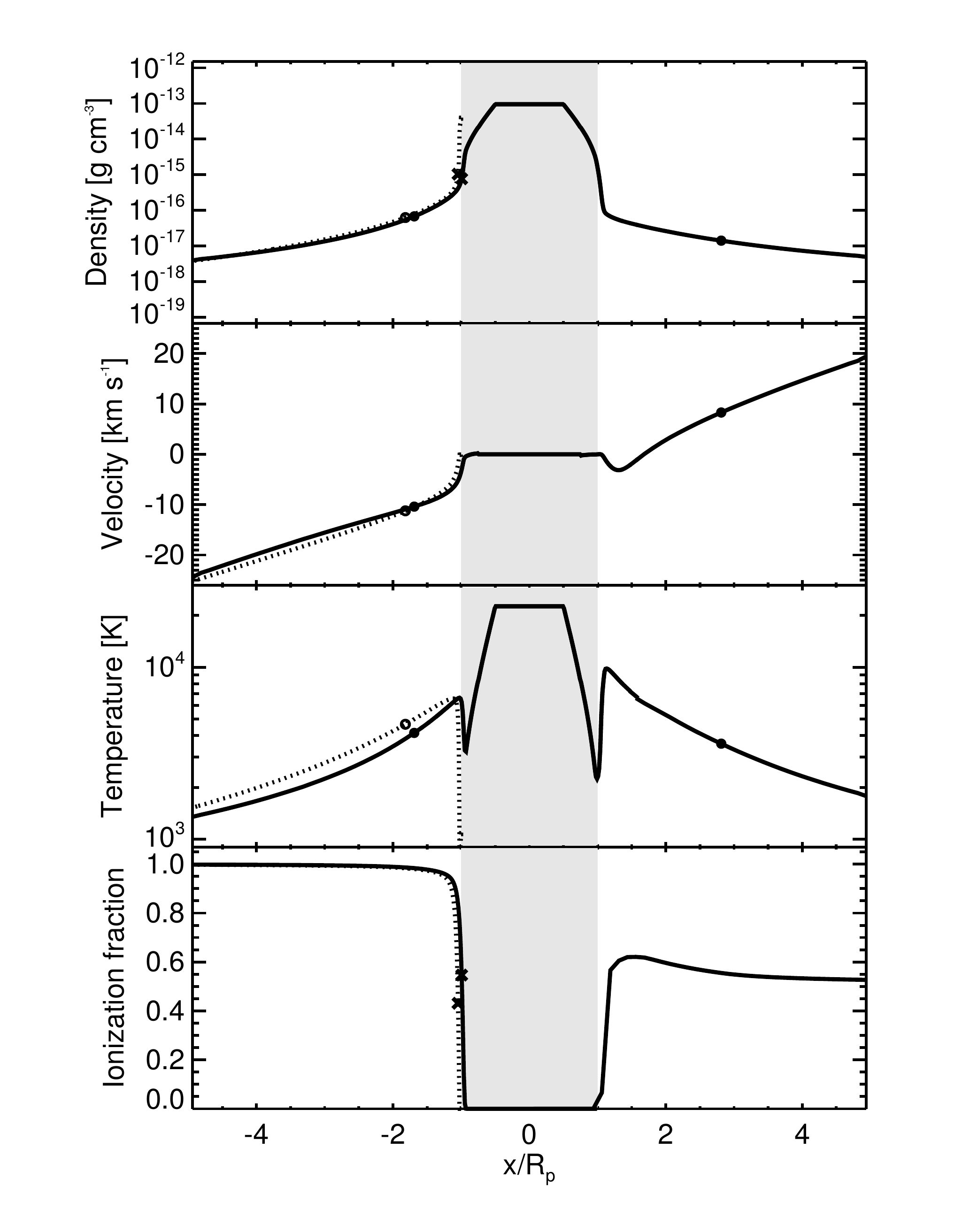}
\caption{Agreement between our model along the substellar ray (solid) and the 1D model of \citet{murray} (dotted), for the same parameter values.  Circles indicate sonic points and $x$'s denote the $\tau=1$ surface to photoionization.  The unshaded regions are those that we model physically. \label{fig:vel1d}}
\end{figure}
We compare a 1D slice from our simulation to the model of MC09 re-calculated for our parameter values.  The 1D model of MC09 is a good point of comparison because it also includes ionization balance, tidal gravity, heating and cooling terms.  While both the day and night sides are available from our 3D model, only the dayside is available from the 1D models.

Figure \ref{fig:vel1d} shows the steady-state density, velocity, temperature, and ionization profiles from both models, along the substellar ray.  We find that the $\tau=1$ surface, denoted by an \textsf{x}, is at $-R_{\rm p}$ in both models.  Above the $\tau=1$ surface, the two wind solutions are comparable.  Both models transition from subsonic to supersonic velocities at similar dayside locations relative to the planet.  We define this sonic point using each model's adiabatic sound speed, rather than isothermal sound speed.  The dayside sonic point is interior to the planet's Roche lobe radius ($2.7R_{\rm p}$).
Wind temperatures from both models agree within 25\%.  The verification of our choice of initial parameters, i.e. $\tau(R_{\rm p})=1$, and agreement with comparable 1D results based on the models of MC09 validates that our model is freely and self-consistently setting its parameters, rather than having fixed boundary conditions.  

For both models, the $\tau=1$ location coincides with the sharp transition in ionization fraction at the wind base.  
In the original MC09 high flux models of a less inflated planet, outflows with such a rapid ionization change are characterized as radiative/recombination-limited.  However, our highly inflated planet's outflow is energy-limited.  
The majority of energy deposited by photoionization is converted into PdV work.  Advection, rather than recombination, balances photoionization at the very base of the wind.  Our planet's highly inflated parameters put its outflow near the boundary of the two regimes.  

\begin{figure}
\includegraphics[trim=0cm 0cm 0cm .7cm, clip=true, width=1\linewidth]{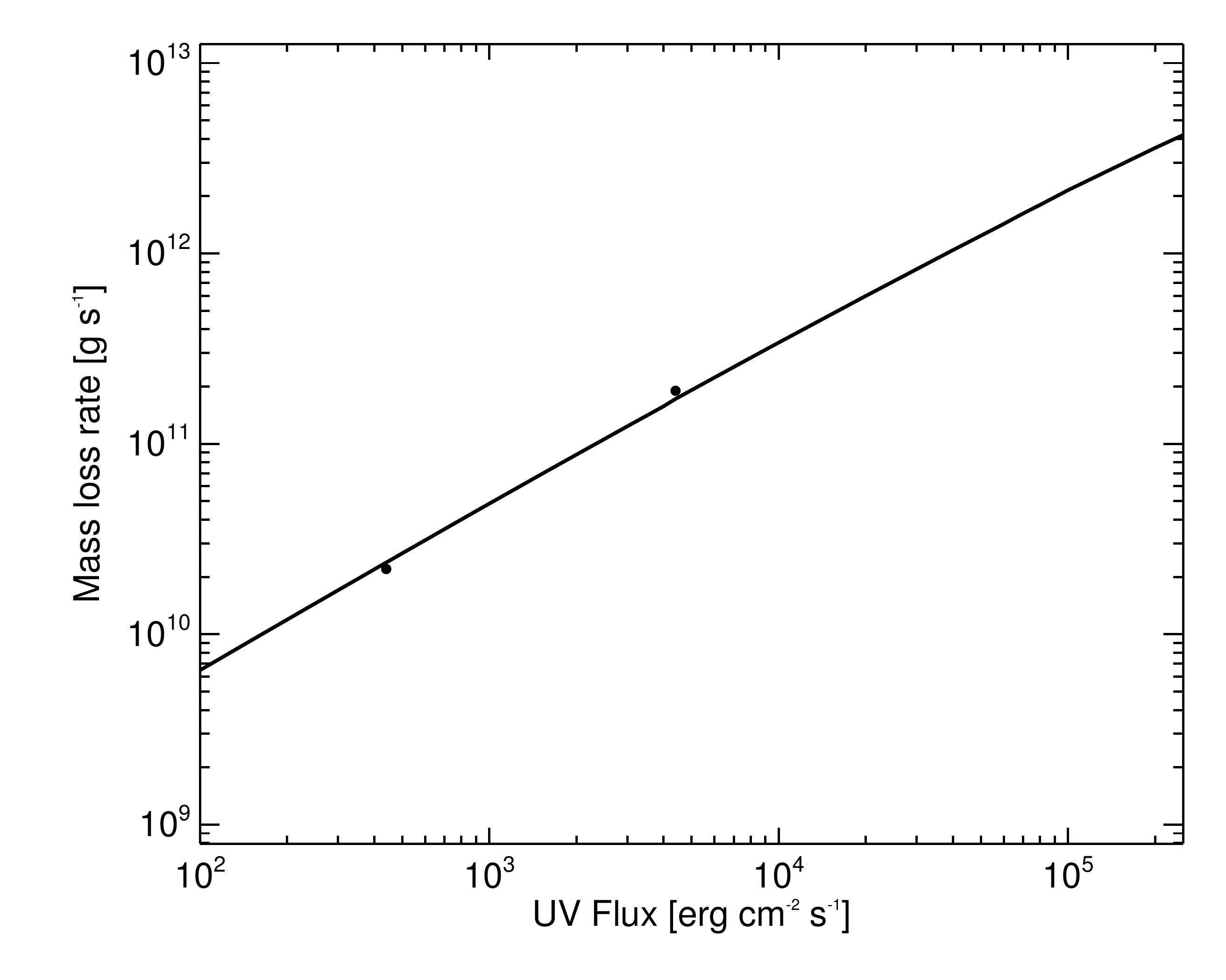}
\caption{Mass loss rates as a function of UV flux from this work (points) and the 1D model of MC09 (solid line), rerun for the parameters used in this work.  The mass loss rate scales as $\dot{M} \propto F_{\mathrm{UV}} ^ {0.9}$ for energy-limited escape.}\label{fig:1dmassloss}
\end{figure}

We compare our mass loss rates to the MC09 1D model, rerun for our stellar and planetary parameters.  We estimate the mass loss rate for each 1D model at a radius $r$ by multiplying the substellar mass loss rate by $4\pi r^2$ and a geometric correction factor, to account for differences in the received flux over the planet's surface.  For energy-limited outflows, MC09 used a constant correction factor of 0.26, derived from mass loss models with different fluxes that showed $M\propto F_{\mathrm{UV}}^{0.9}$.  

Our mass loss rates for maximum photon fluxes of $2.02\times10^{14}$ and $2.02\times10^{13}$~cm$^{-2}$~s$^{-1}$, respectively, agree with the 1D model loss rates and UV flux scaling, as shown in Figure \ref{fig:1dmassloss}.
We are unable to make direct comparisons to other mass loss predictions, due to our choice of model parameters.  However, given that MC09 had mass loss rates in good agreement with other 1D models for HD 209458 \citep{yelle, tian, garciamunoz}, our agreement with the MC09 model recalculated for our parameters suggests agreement between ours and other previous 1D models. 

\section{Predicted Lyman-alpha extinction}\label{sec:obs}
\begin{figure}
\centering
\includegraphics[trim= 0.3cm 0cm 0cm 0cm, clip=true, width=1 \linewidth]{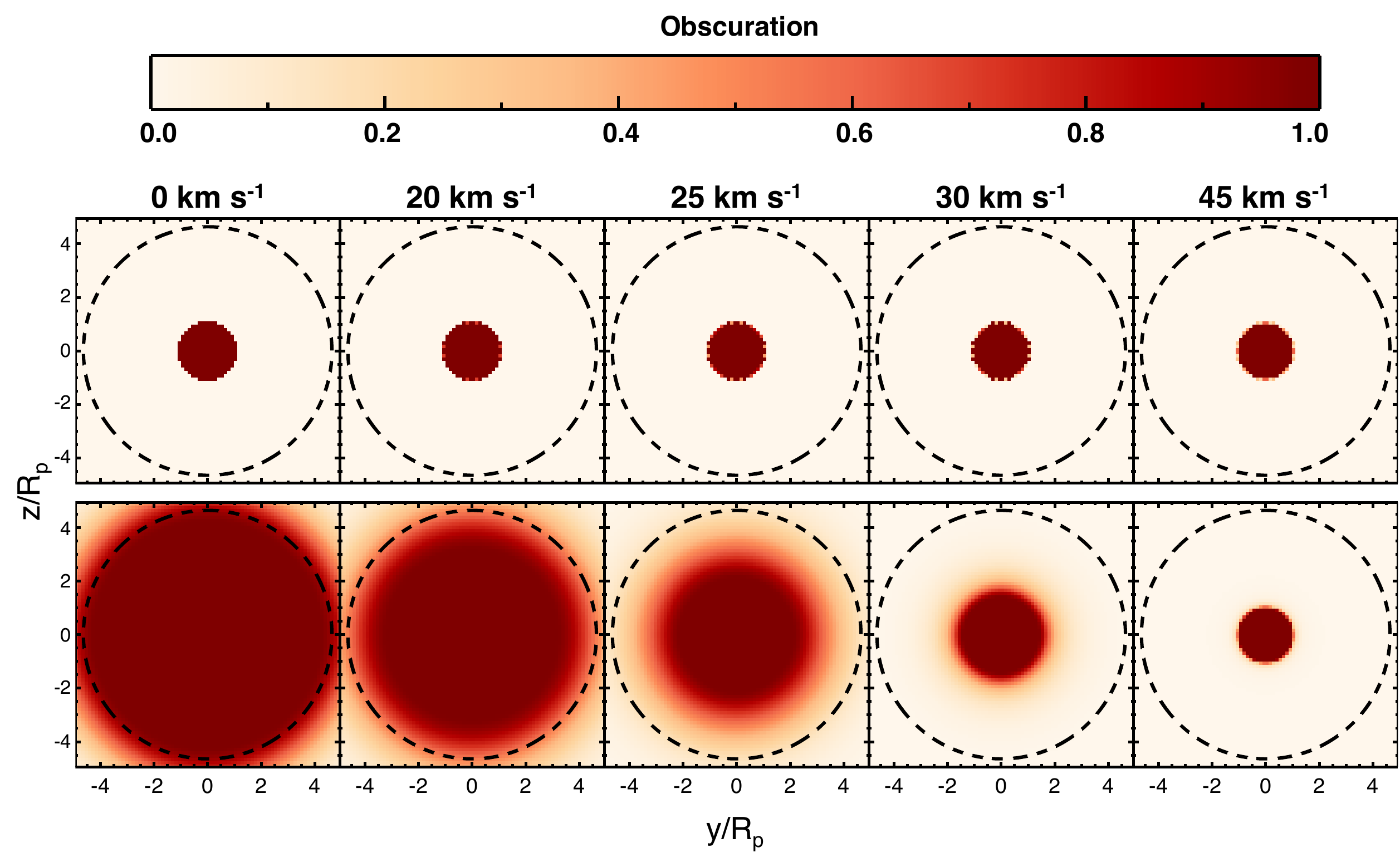}
\caption[]{Channel maps of the spatial distribution of~\Lya~obscuration at line center and then off-center by the specified velocities.  The upper panel shows the initial obscuration and the lower panel shows steady-state outflow obscuration, at $1.39\times 10^6$s.  The dashed circle shows the spatial extent of the stellar disk.  \label{fig:cmap}}
\end{figure}

\begin{figure}
\includegraphics[trim= 0cm 0cm 0cm 0cm, clip=true, width=1\linewidth]{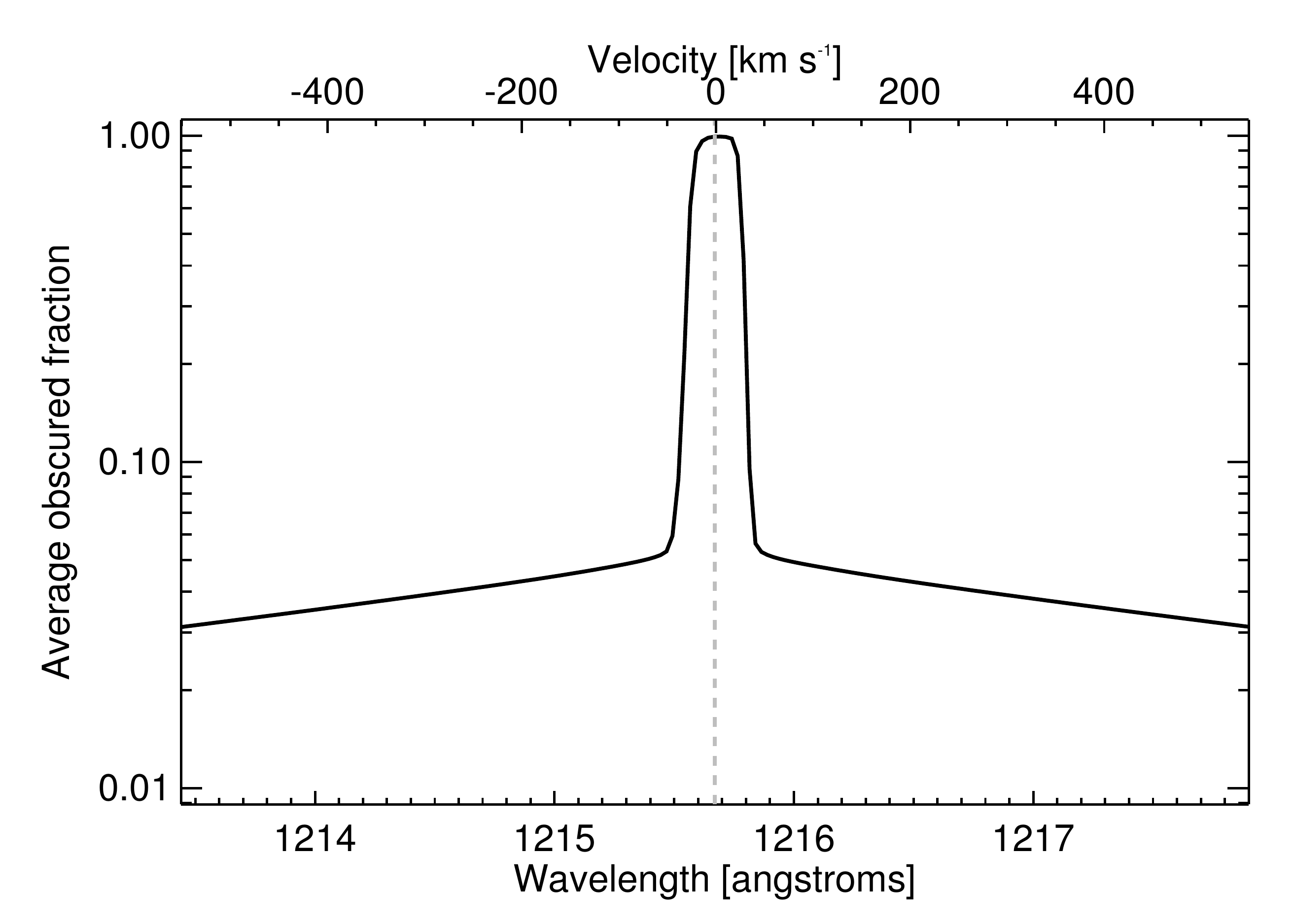}
\caption{Average \Lya~obscuration integrated over the stellar disk, caused by steady-state outflows.  \label{fig:aveobs}}
\end{figure}

For outflows from our pure hydrogen planet, we calculate the predicted obscuration of stellar \Lya~emission during transit.  We focus on \Lya~because it has been observed for several planets including HD 209458b, HD 189733b, and GJ 436b.  Our calculation serves as a first step toward motivating future observations of a wider range of exoplanets, including WASP-17b.

To determine the spatial extent of \Lya~absorption, we use the temperature and ionization fraction of each cell in our 3D simulation to compute a Voigt line profile, Doppler shifted by the cell's bulk radial velocity.  We use values from \citet{morton} for the \Lya~oscillator strength, Einstein $A$ coefficient, and line center wavelength.  In this work, we calculate the obscuration produced by a planet passing across the stellar disk mid-transit.  We assume that the star has $R_{\star} = R_{\odot}$.  

Given that our simulation is in 3D, we can examine the integrated absorption spatially, in 2D.  Channel maps of the obscuration, $1- \exp (-\tau)$, initially and in steady-state, are shown in Figure \ref{fig:cmap}.  Extinction from the outflow is axisymmetric across our simulated volume.  The velocity dependence of the extinction is shown in these maps and more clearly seen in Figure \ref{fig:aveobs}, which shows the total obscuration averaged over cells in the stellar disk.  To make this figure consistent with traditional sign conventions, the velocity signs have been reversed and are thus the opposite of other velocity values in this paper.  The obscuration is roughly symmetric about line center and drops to 5\% at $\pm \approx 50$~km~s$^{-1}$.  At higher velocities, in the line wings, obscuration of a few percent comes only from the planetary disk.  Confounding geocoronal emission and absorption by the interstellar medium make observations of \Lya~absorption challenging near line center, where wind obscuration is most prominent.

\section{Summary and Discussion}
 \label{sec:discussion}
We have simulated global hot Jupiter outflows in three-dimensions (3D).  The outflow is self-consistently launched by photoionization heating, which we simulate using a new implementation of  planar radiative transfer in the Athena hydrodynamic code.  Our code is compatible with multiple levels of static mesh refinement (SMR), distributed across parallel processors, and is freely available for further use.  

We drive outflows from a highly inflated hydrogen planet, illuminated by a plane-parallel source of ionizing radiation.  We find that photoionization heated supersonic outflows emerge on the dayside and advect around the planet, launching nightside outflows not captured in earlier 1D simulations.  Outflows are everywhere ionized, except in the planet's nightside shadow, where the outflow is largely neutral.  On the planet's night side, a stagnation point in the flow separates outflowing gas from a time-variable circulation region.  Perhaps counterintuitively, the longer residence time of gas in this region allows it to reach temperatures higher than those achieved on the planet's dayside.  For our fiducial parameters, gas within the nightside stagnation region is the only portion of the flow that is substantially radiatively cooled.

For the highly inflated planet we consider, we find mass loss rates of $1.9 \times 10^{11}$~g~s$^{-1}$ and $2.2\times10^{10}$g~s~$^{-1}$ for photon fluxes of $2.02\times10^{14}$~cm$^{-2}$~s$^{-1}$ and $2.02\times10^{13}$~cm$^{-2}$~s$^{-1}$, respectively.  The outflow is marginally energy-limited.  Our mass loss rates are consistent with the 1D escape model of MC09, rerun for the same parameters.  Along the dayside substellar ray, we also find remarkable agreement between the outflow structure in our simulations and the 1D model.

A benefit of our 3D model is the ability to examine not only day and night differences in outflows but also position-dependent extinction.  Neutral, shadowed gas, is a major contributor to the absorption of stellar \Lya~emission - a key predicted observable.  Integrating through the box, we find that the \Lya~absorption of the escaping gas obscures stellar emission out to 
$ \pm \approx 50$~km~s$^{-1}$.
 
 The work presented in this paper is the starting point for more realistic modeling of atmospheric escape.  While we have simulated asymmetries in stellar heating, the outflow will also be shaped by the Coriolis force, magnetic fields, and interactions with the stellar wind.  In addition, true atmospheres are irradiated by a full spectrum of energetic photons.  As our work is motivated by observations of mass loss from close-in planets, in the following paragraphs we review how these physical effects may change the observed transmission spectrum.  Since \Lya~observations do not yet exist for our planet or the similar WASP-17, we discuss the impact of these physical processes in the context of the observed high velocity Lya~obscuration at $\pm 100$km s$^{-1}$ for HD 209458b \citep{vidal03}.  Bear in mind that the two planets have different parameters and thus will not have identical transmission spectra, even with additional physics.

To distinguish 3D geometric effects on the outflow, we have used a monochromatic flux source and compared our results to 1D models, which also used monochromatic flux sources.  Energetic effects involved in using a full stellar spectrum were not studied in this first 3D photoionization study.  Due to the wavelength dependence of the photoionization cross section, $\sigma_{\rm ph}$, a full stellar spectrum can smear out the $\tau=1$ surface where photons are absorbed, increasing the thickness of the wind base \citep{trammell, koskinen13, koskinen}.
   These changes may increase the column density near the base of the wind or increase the radial extent of neutral hydrogen \citep{koskinen13}.  These effects can contribute to a broadened line profile.  At very high incident fluxes, photoionizing X-rays, which have a lower photoionization cross section per hydrogen nucleus and hence are deposited deeper in the atmosphere, can inject heat quickly enough to be a dominant driver of the outflow \citep{owenjackson}.  The resulting wind structure is analogous to that modeled for UV-driven winds but begins deeper in the atmosphere, increasing the total wind column.  
To study this effect, additional spectral bins including photoionization by X-rays can be included in future versions of our model.

High velocity neutral gas may arise in a bow shock that marks the interaction between the outflow and the stellar wind.  \citet{tremblin} find that charge exchange between outflowing neutral atoms and faster stellar wind ions in this wind-wind interaction region can produce a sufficient fast neutral population.  

\citet{trammell,trammell14} have suggested that if the planet's magnetic field is dipolar, a magnetically confined dead zone near the equator does not participate in the outflow and could host an enhanced column of neutral gas.  MC09 estimate that if the line-of-sight column of  neutral gas were enhanced by a factor of 3--5, then  naturally broadened line wings from this slower-moving population would be sufficient to produce the observed obscuration (c.f. Figure \ref{fig:aveobs}; see also \citealt{benjaffel}). 
Regardless of whether magnetic confinement enhances the neutral column,  \citet{adams} demonstrates that for reasonable hot Jupiter magnetic field strengths, the magnetic field will redirect and possibly reduce the atmospheric outflow, affecting its 3D structure.  For our fiducial model, we find that a local  field strength $B \gtrsim 0.06$G is sufficient for the magnetic pressure $B^2/8\pi$ to overcome the ram pressure of the wind at all radii.  For example, at the sonic point of the wind located at $\sim$2$R_{\rm p}$, this local $B$ corresponds to that from a dipole field with a surface strength of $0.5$G.   \citet{owenadams} find using 2D models that  increasing magnetic field strength can suppress nightside outflows and reduce planetary mass loss rates. 
Because we have used Athena, an MHD code, for our simulations, it can be extended to study the effects of the magnetic field on planetary wind confinement.  

A line-of-sight column density enhancement could alternatively result from the confinement of the atmospheric outflow by the stellar wind.  Depending on the level of stellar activity,  pressure confinement by the stellar wind can channel outflowing gas into a smaller solid angle, increasing its local density \citep{proga}.  In extreme cases, the flow is confined to a cylindrical region on the planet's night side.  The Coriolis force (due to the planet's orbital motion) will limit the extent of this high-density region by  causing the outflow to curve.  
The Coriolis force bends trajectories on a timescale $\sim\Omega^{-1}$, where $\Omega$ is the angular frequency of the planet's orbit.  For our outflow velocity of $v \sim 20$ km s$^{-1}$, 
the length scale of this curvature is 
$L_{\rm curve} \sim v\Omega^{-1} \sim 10^{11}$ cm, about twice the Roche lobe radius.  

Finally, we note that the Coriolis force will produce curvature in the neutral shadow highlighted here and in the 2D model of \cite{owenadams}.  Without such curvature, this large reservoir of neutral gas does not contribute to the line-of-sight absorption signature since the planet lies between it and the star.  At distances larger than $L_{\rm curve}$, the neutral gas will curve out of the planet's shadow, where it has the potential to substantially enhance the transit absorption signature on the side of the planet opposite to its direction of motion.  Upon leaving the nightside, a neutral atom will ionize within $\sim (F\sigma_{\rm ph})^{-1} \sim$ a few hours, sufficient time to travel approximately a planetary radius at 20 km s$^{-1}$.

Incorporating the Coriolis force into our simulation will clarify how stellar photoionization changes the outflow structure and allow us to produce more realistic, spatially resolved extinction maps and spectra.  With the introduction of a stellar wind and magnetic fields, comparisons can be made to 3D studies of these effects that did not include self-consistent photoionization driven wind launching  \citep{cohen, trammell, matsakos} .  

While the model itself can be expanded to include additional physics, post-processing will also offer avenues for other comparisons to observation.  Our hydrogen results can be translated into other species, which can be used to look for observational signatures of outflows.  Given the 3D nature of our code and the ability to track the time evolution of the gas, we can use a larger computational box to look for sightline specific effects that would appear at ingress and egress.  Such a model would allow us to make comparisons to observations with sufficient time resolution, including \citet{kulow} and \citet{lecavelier12}.  With improved physics and simulated observational output, we will be one step closer to motivating and understanding observations of mass loss from hot Jupiters.

\acknowledgements
We thank Xuening Bai, Jim Stone, and James Owen for helpful discussions.  We are grateful to the referee for helpful comments that improved the paper.  A.T. was supported by the National Science Foundation Graduate Research Fellowship under grant no. DGE-1144152. K.M.K. was supported in part by NASA through Hubble Fellowship grant no. HF-51306.01 awarded by the Space Telescope Science Institute, which is operated by the Association of Universities for Research in Astronomy, Inc., for NASA, under contract NAS 5-26555.  R.A.M. acknowledges support from NSF grant AST-1411536.  M.R.K. acknowledges support from NSF grant AST-0955300.  The computations in this paper were run on the Smithsonian Institution High Performance Cluster (SI/HPC) and the Odyssey cluster supported by the FAS Science Division Research Computing Group at Harvard University.

\appendix
{
\section{Ionizing Radiative Transfer Setup and Validation}
 \label{sec:appendix}
\subsection{Implementation with SMR and Parallelization}
We have extended the ray-tracing radiative transfer method from \citet{krumholz07f} to Athena grids with multiple levels of static mesh refinement (SMR), parallelized with MPI.  We use a plane-parallel source of ionizing radiation, with rays aligned with the grid.  As a result, we loop over cells along the direction of radiation propagation and iteratively carry out radiation updates.  Radiative source terms, given on the right hand side of  Equations \ref{eqn:energy} and \ref{eqn:advection}, are calculated and used to update the neutral fraction and energy density, using the operator-split approach described in \citet{krumholz07f}.  Key to calculating the photoionization source terms is the ionizing flux in each cell.  For rays passing through different levels of resolution, we communicate ionizing fluxes across each level and processor boundary.

For a given timestep, our steps for ionizing radiative transfer on a grid with multiple SMR levels, from coarsest (level 0) to finest resolution are:
\begin{enumerate}
\item Integrate the flux along rays on level 0.  
\item Compute ionization, heating, and cooling rates.  Update the energy density and neutral fraction in each cell. 
\item Evaluate stopping criteria on level 0, and store the iteration timestep.
\item Prolongate the flux from the current, coarse level (e.g. level 0) onto the next level of resolution (e.g. level 1), where the coarse grid first crosses into the finer grid.  To prolongate the flux, copy the coarse flux into the overlapping fine cells, so that each coarse level ray sub-divides into eight rays with equal flux, at the next level of resolution.  

\item Integrate the flux along the next level of resolution (e.g. level 1), and repeat step 2. 
\item Continue this level's integration and radiation update until the coarsest, level 0, timestep has been reached.  
\item Repeat steps 4-6 for the remaining resolution levels.

\item Restrict the neutral fraction and energy density from finer, higher resolution levels to coarser levels, e.g. level 2 to level 1, level 1 to level 0, by averaging the finer values of the eight overlapping cells to obtain one value for the coarse level ray.
\end{enumerate} 
  
As described above, when SMR is used, the coarsest grid level is evaluated first, followed by subsequently finer levels.  For each level, we integrate the flux at each cell using Equation \ref{eqn:flux} with a discrete calculation of the optical depth,  Equation \ref{eqn:optdepth}.  Calculated fluxes are used to compute radiative source terms and update the neutral fraction and energy density in each cell.  As these radiation updates are computationally cheap relative to hydrodynamic updates, multiple radiation updates can be carried out before entering the hydrodynamics integrator.  We use the stopping criteria of \citet{krumholz07f} to determine when the coarse grid radiation update should terminate.  To ensure that propagation speeds are the same across different resolution levels, we require finer levels to complete radiation updates for the same elapsed time as that of the coarse level.

With SMR, the flux is needed at the boundaries between different levels of resolution, so we prolongate the radiative flux from lower resolution levels to cells in overlapping higher level cells.  Resolution level boundaries and overlapping regions are determined during the simulation initialization.  Prolongation occurs at every time step, only at level boundaries, since the flux is then iteratively calculated along grid-aligned rays.  Photon number is conserved across level boundaries, since we copy the radiative flux in each coarse cell into the finer cells, which have smaller areas.  Recall that Athena requires each level of resolution to be smaller by a factor of 2, in each dimension.  

Following the radiation update, we do not restrict the radiative flux from the fine grids back to the coarse grids.  In our mass loss simulation, the ambient gas is optically thin and the planet is optically thick, so any flux that needs attenuating will be visible on the coarse levels as needed.  We do, however, include a restriction step from the finest levels back to the coarse levels, to recombine finer rays together, for the hydrodynamic variables, $E$ and $\rho_n$.

When using parallelization, extra communication is required to coordinate, send, and receive ionizing fluxes on different processors.  SMR prolongation and restriction operations are complicated by the fact that values are distributed on different processors.  To appropriately direct fluxes to the grid levels spread across multiple processors, we store and communicate processor identifiers for the underlying coarse and overlapping fine grids.  With this information, flux prolongation now occurs as two separate steps - a sending of the flux values from the coarse grid and a receipt of the values on the fine grid.  

Although parallelization speeds up our simulation, there are restrictions.  For our plane parallel attenuation of the flux, grids further downstream in the direction of radiation propagation must wait for integration to happen upstream.  While radiation updates are computed on the grids in order along the direction of radiation propagation, there are no barriers to carrying out radiation updates on the grids distributed orthogonally to the radiation propagation.  As previously noted, we also require coarse levels to complete their radiation updates before finer, overlapping regions can.  
\begin{figure}
\includegraphics[width= 1 \linewidth]{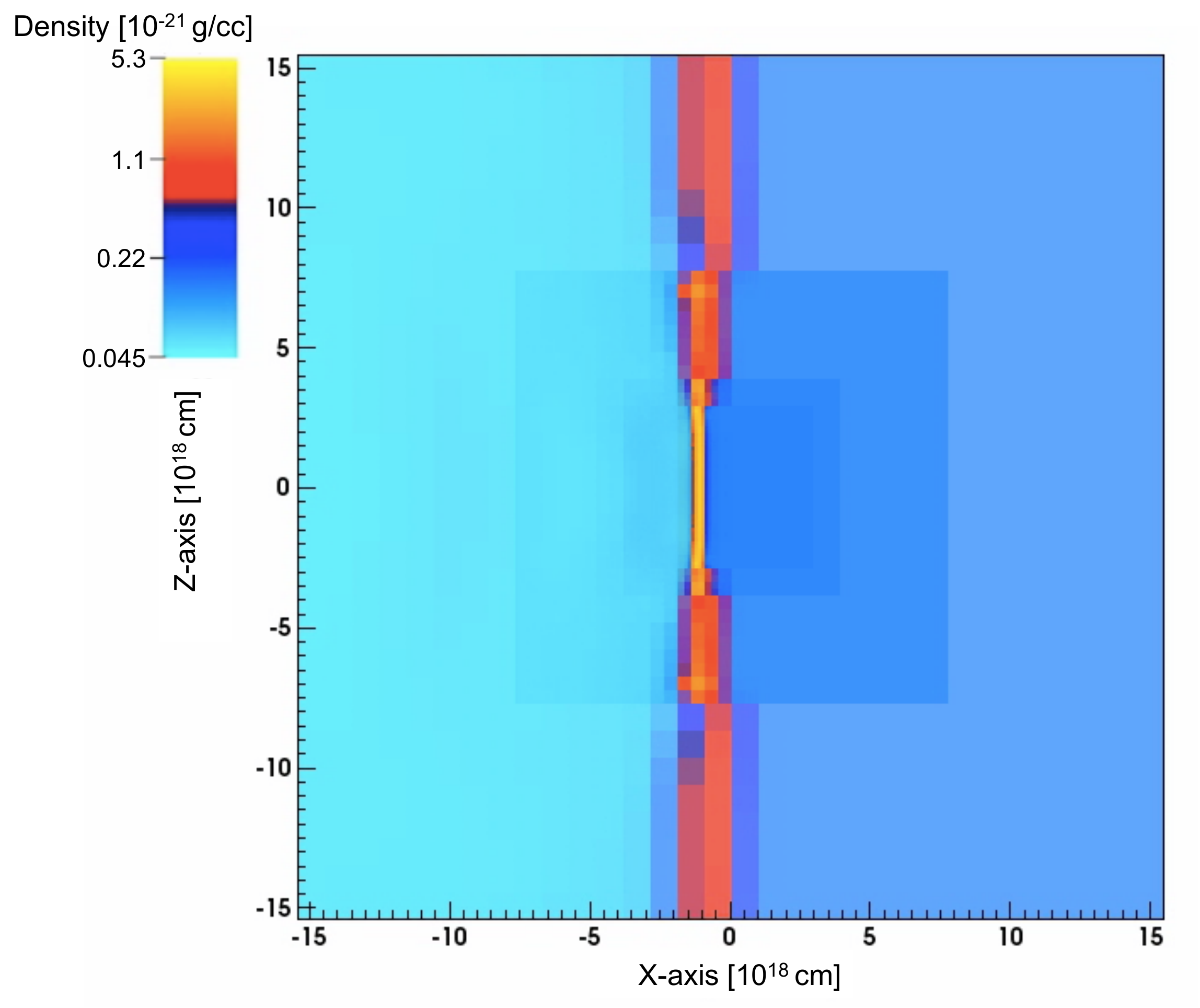}
\caption{2D midplane slice showing the density of an ionization front in a uniformly dense gas, simulated with four levels of resolution.  To highlight resolution level boundaries, densities in this slice have been rendered with different levels of transparency for each resolution level - coarse levels are more transparent.  Radiation is incident from the -$x$ direction.  
\label{fig:frontsmr}}
\end{figure}
\subsection{Implementation Validation}
 \label{sec:accuracy}

We examine the accuracy of our radiation module with SMR and parallelization, by comparing our code to analytic expectations for a planar ionizing source incident on a uniform, optically thick neutral medium.  The evolution of the ionized region is akin to that of an HII region Str\"omgren sphere, but with a planar geometry.

We set up a (50~pc)$^3$ box of neutral gas ionized by a planar source with a photon flux of $3.67\times10^9$~cm$^{-2}$~s$^{-1}$, as in \citet{gendelev}.  To test our implementation, we use a grid with four levels of resolution, distributed among 8 processors, with each processor hosting four grids - one for each level of resolution.  All resolution levels were centered within the box, with the coarsest three levels each having 32 cells, and the finest resolution level having 48 cells.  To appropriately make comparisons with an HII region, we initialize the gas to an ISM appropriate chemistry.  Considering atomic hydrogen with helium, we use the mean gas mass per hydrogen nucleus of $\mu_{\rm H} = 2.3\times10^{-24}$~g and the mean mass, $\mu = 2.1\times10^{-24}$~g.  We use a carbon abundance of $\alpha_C = 3\times10^{-3}$ \citep{sofia}.  We initialize the gas to a uniform neutral medium with number density $n_0$ = 63 cm$^{-3}$ and an isothermal sound speed $c_{s,0}=5.4\times10^4$~cm~s$^{-1}$.  Unlike our hydrogen mass loss simulation, we use $n_{\rm e} = n_{{\rm H}^+} + \rho \alpha_C / (14 \mu_{\rm H})$, where $\alpha_C$ is the carbon abundance of the gas.  Since neutral carbon has an ionization threshold below 13.6 eV, C$^+$ is the dominant ionization state in the neutral ISM, and this provides a minimum abundance of free electrons.
 
For HII regions, different radiative source terms are needed in the fluid equations.  We include all terms from Sections \ref{sec:ioniz} and \ref{sec:heating} except for \Lya~line cooling.  Additionally, we add contributions from collisional ionization to our ionization balance, using the rate per unit volume of \citet{tenorio}
\be
\mathcal{I_{\rm coll}} \approx \left(5.84 \times 10^{-11}{\rm cm}^3~{\rm s}^{-1}\right) n_{\rm e} \rho_n \sqrt{T\over{\rm K}}e^{(-13.6{\rm ~eV})/kT}.
\ee
We also include the optically thin cooling terms for molecular cooling, line cooling from O, N, and Ne, and free-free cooling, as described in \citet{krumholz07f}.  A full inclusion of heating and cooling terms is necessary to thermostat the ionized gas to $10^4$~K. 

We can qualitatively assess the success of our radiative transfer implementation by looking at a 2D slice of the gas density, shown in Figure \ref{fig:frontsmr}.  Note that to make the different SMR regions visible, we rendered the product of the density and a constant transparency value, with greater transparency values for coarse resolutions.  Radiation incident from the left edge of the box photoionizes the gas, increasing its temperature and lowering its density.  The ionized region is separated from the neutral region by an ionization front (red), which we resolve to one grid cell.  As a result, the front appears thinner in the higher resolution regions, and it is also denser - to conserve mass.   The ionization structure is more clearly seen in Figure \ref{fig:frontstruc}, which gives a 1D view of the density, temperature, neutral density, and ionization fraction.  The ionization front is visible as the peak in density, separating the lower density ionized region from the higher density neutral background.

Quantitatively, we can compare our simulation output to analytic calculations of ionization front expansion in HII regions.  When an ionization source first turns on, the initial rate of photoionization overwhelms the rate of recombination.  The result is a rapid phase of ionization and expansion, faster than the local sound speed.  As this R-type expansion slows down due to reaching near ionization equilibrium at the Str\"omgren length, D-type expansion occurs, driven by an overpressure in the photoionized region.  At times much greater than the sound crossing time of the ionized region, the photoionized gas will reach uniform pressure and density and exert a pressure on the swept up front of neutral material that bounds the ionized region.  We can use the requirement from momentum conservation, that the photoionized region's force on the neutral gas be equal to the neutral gas's rate of momentum change per unit area, to derive the size of the ionized region.  The ionized region's pressure on the swept up, neutral gas is
\be
P_{\rm i} =  c_{s,{\rm i}}^2 \rho_{\rm i}= c_{s,{\rm i}}^2 \mu_{\rm i} n_{\rm e} =  c_{s,{\rm i}}^2 \mu_{\rm i}  \sqrt{\frac{F_0}{\alpha_B l}},
 \label{eqn:p0}
\ee
where $l$ is the size of the photoionized region, $c_{s,{\rm i}}$ is the sound speed in the photoionized region and $\mu_{\rm i}=\mu_{\rm H}/2$ is the mean gas mass of the ions.  We assume that the electron number density $n_{\rm e}$ equals the ion number density $n_i$ because the photoionized region is in ionization equilibrium.  At late times, the gas will be in thermal equilibrium, so we assume that $c_{s,{\rm i}}$ is a constant.  The time rate of change of the neutral material's momentum is 
\be
 P_0 ={1\over A}\frac{d}{dt}(mv) = \frac{1}{A}\frac{d}{dt} ( n_0 \mu A \Delta l\dot{l})\approx \frac{d}{dt} ( n_0 \mu l\dot{l}) ,
 \label{eqn:p1}
 \ee
where $A$ is the cross-sectional area of the front and we have assumed that the distance over which mass has been swept, $\Delta l$, is comparable to the size $l$ of the ionized region.  Equating Equations \ref{eqn:p0} and \ref{eqn:p1} and assuming a similarity solution of the form $l\propto t^n$  yields
\be
l(t) = C't^{4/5}, ~~C' = \left( \frac{25}{12}\sqrt{\frac{F}{\alpha_B}} \frac{c_{s,{\rm i}}^2\mu_{\rm i}} {n_0 \mu}  \right) ^{2/5}.
\label{eqn:ifrontanalytic}
\ee

\begin{figure}
\centering
\subfigure{
\includegraphics[width=1\linewidth]{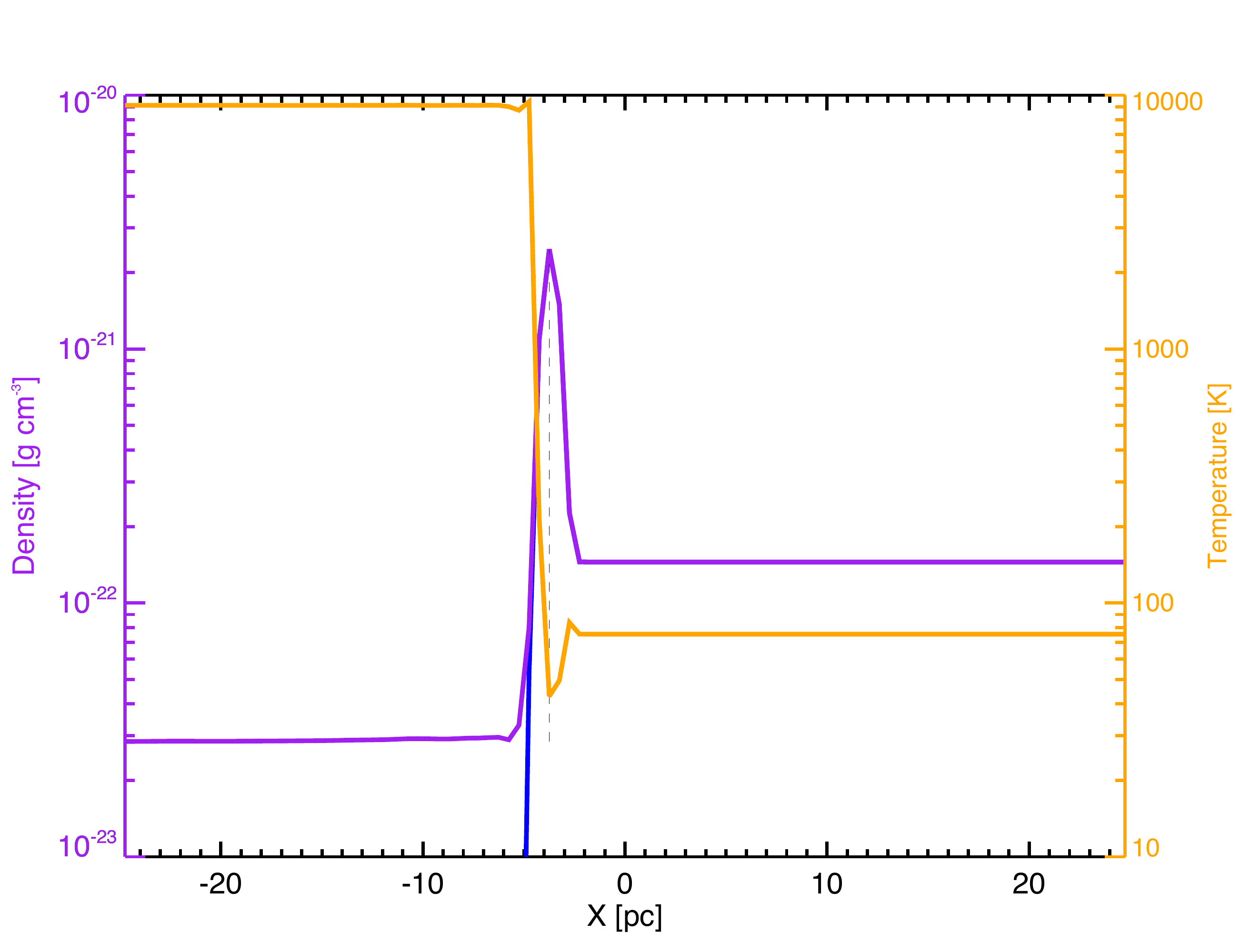}
}\\
\subfigure{
\includegraphics[width=1\linewidth]{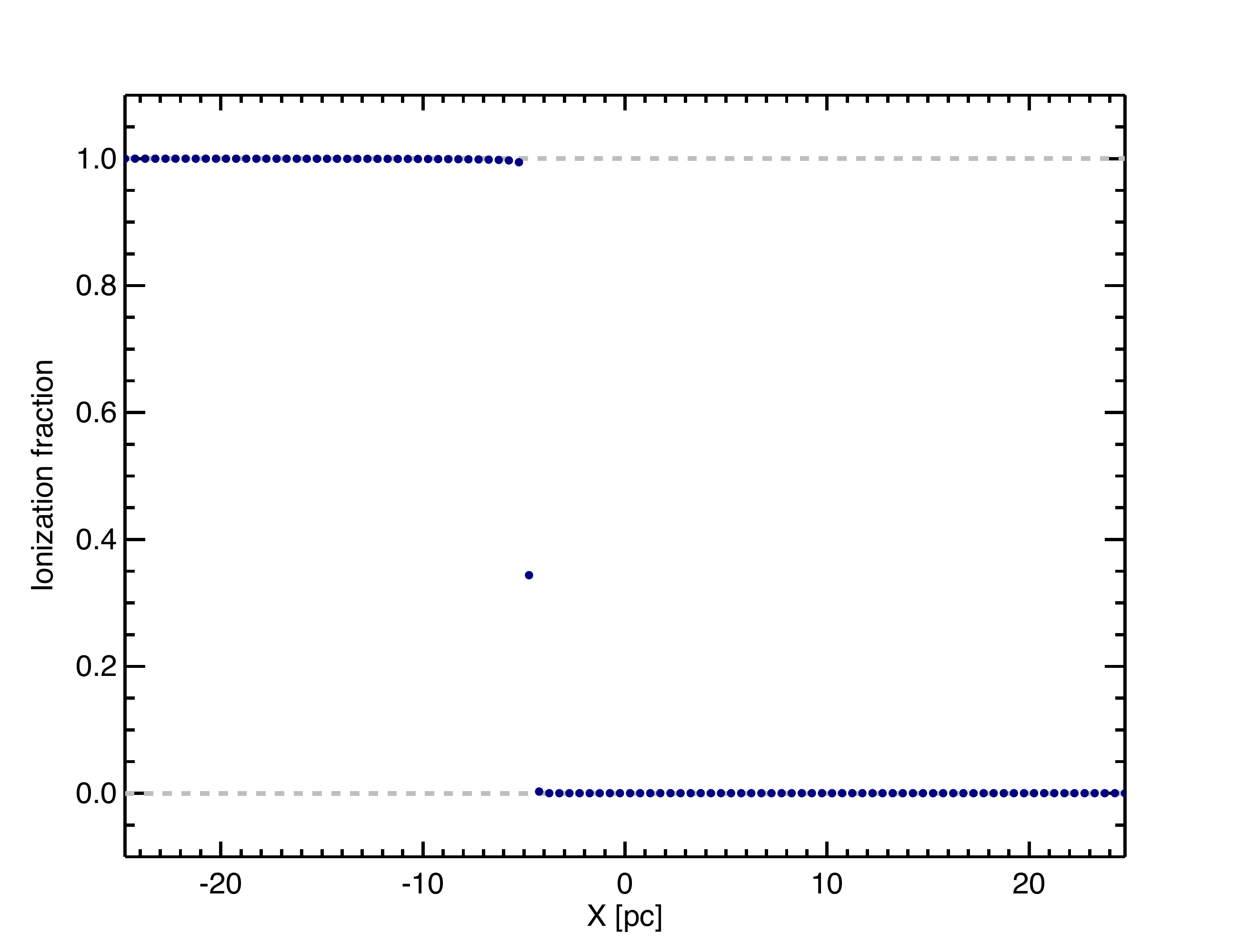}
}
\caption{Ionization front structure at $t=3.8$~Myr, with total density (purple), neutral density (blue), and temperature (orange), shown in the upper plot.  The lower plot shows the ionization fraction.  The dashed line in the upper plot is the front location, used for Figure \ref{fig:frontloc}.  To the right of the front location, the neutral density is equal to the total density  and thus hidden. \label{fig:frontstruc}}  
\end{figure}

\begin{figure}
\includegraphics[width=1\linewidth]{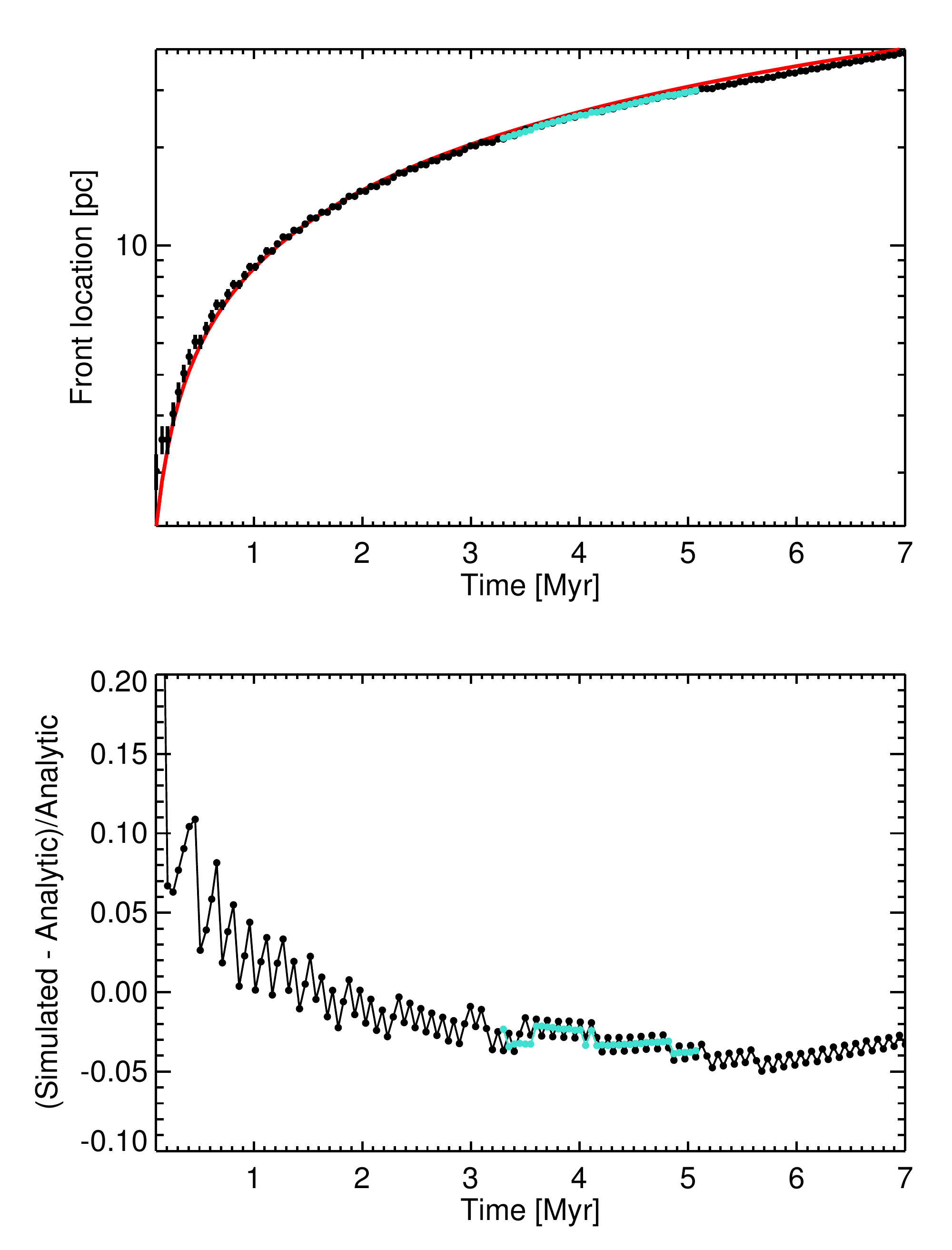}
\caption{Comparison of the ionization front location for a coarse domain (black), with an SMR region (blue), to D-type analytic approximation (red) as a function of time (top), with residuals (bottom).  \label{fig:frontloc}}
\end{figure}

Figure \ref{fig:frontloc} shows the agreement between our simulation and analytic approximation for a coarse (black) and finer SMR region (blue).  For this plot, the front location, or size of the ionized region corresponds to the $x$-location of maximum density, for a slice through the box at fixed $y$ and $z$; an example of this location is highlighted by the dashed line in Figure \ref{fig:frontstruc}.  Choice of other slices orthogonal to the radiation direction yields similar results, as the gas is symmetric in both of these directions.  We determine the corresponding analytic front location, using the simulated ionized sound speed as input into Equation \ref{eqn:ifrontanalytic}.  The result we find is that the front location agrees with the analytic expression to within 5\%.  The oscillatory behavior of the residuals is a result of output that is not commensurate with the propagation time across one cell.  Nevertheless, we find that the coarse and fine regions have consistent agreement with the analytic approximation.  While we have tested the physical validity of our radiative transfer algorithm with SMR and MPI, we have not examined its efficiency and scaling.

\section{Convergence of Mass Loss Simulations}\label{sec:converge}
\begin{figure}
\includegraphics[trim= 0cm 0cm 0cm 0cm, clip=true, width=1\linewidth]{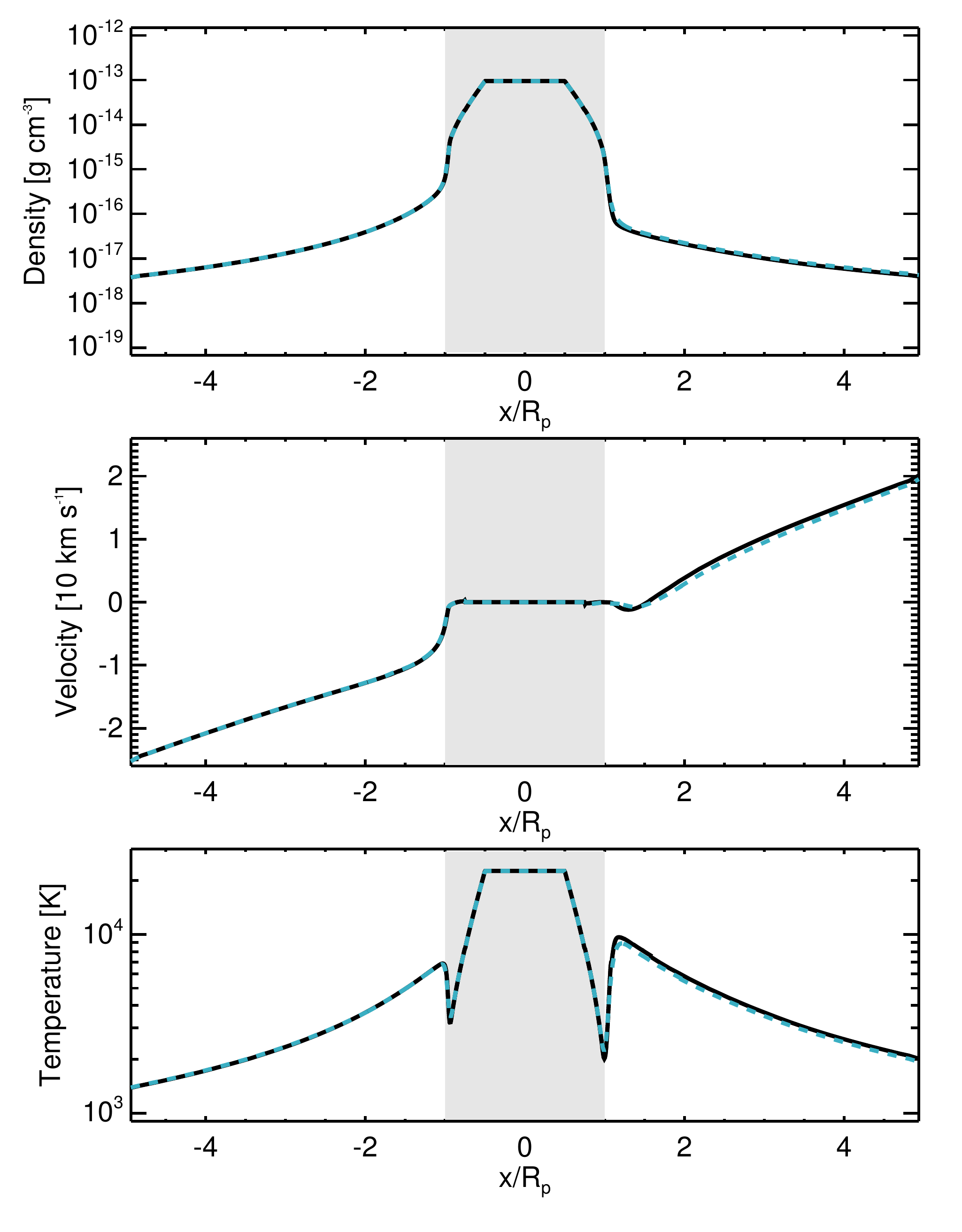}
\caption{High resolution (solid) and lower resolution (dashed blue) simulations have well-converged outflow results, as shown in these 1D profiles.  \label{fig:convergence}}
\end{figure}
To validate our numerical results, we run a convergence test by decreasing the resolution of the planet's atmosphere by a factor of two.  Since additional refinement levels increase the resolution of our planet, not the entire box, we re-run our fiducial simulation with four levels of refinement, instead of five.

As highlighted in the 1D profiles in Figure \ref{fig:convergence}, these two simulations are converged and produce comparable outflows.  As expected, differences arise only in the stability of the static atmosphere, since lower resolution decreased the number of resolved scale lengths.  Thus, the only differences in this resolution test were seen around the edge of the atmosphere, not in the escaping gas.  We find no difference in the calculated mass loss rates for the two resolutions, further suggesting that our results are converged.  

In the main text, we present our high resolution results for our fiducial parameters.  Due to the convergence we describe here, from our resolution tests of four and five levels of refinement, we use only four levels of refinement for the low-flux simulation described in the main text.


\begin{thebibliography}{}

\bibitem[Adams(2011)]{adams} Adams, F.~C.\ 2011, \apj, 730, 27

\bibitem[Ballester \& Ben-Jaffel(2015)]{ballester} Ballester, G.~E., \& Ben-Jaffel, L.\ 2015, \apj, 804, 116 

\bibitem[Baraffe et al.(2004)]{baraffe} Baraffe, I., Selsis, F., Chabrier, G., et al.\ 2004, \aap, 419, L13 

\bibitem[Ben-Jaffel(2008)]{benjaffel} Ben-Jaffel, L.\ 2008, \apj, 688, 1352

\bibitem[Bento et al.(2014)]{bento} Bento, J., Wheatley, P.~J., Copperwheat, C.~M., et al.\ 2014, \mnras, 437, 1511 

\bibitem[Bisikalo et al.(2013)]{bisikalo} Bisikalo, D., Kaygorodov, P., Ionov, D., et al.\ 2013, \apj, 764, 19 

\bibitem[Black(1981)]{black} Black, J.~H.\ 1981, \mnras, 197, 553 

\bibitem[Bourrier \& Lecavelier des Etangs(2013)]{bourrier} Bourrier, V., \& Lecavelier des Etangs, A.\ 2013, \aap, 557, A124 

\bibitem[Chamberlain(1963)]{chamberlain} Chamberlain, J.~W.\ 1963, \planss, 11, 901 

\bibitem[Charbonneau et al.(2000)]{charbonneau} Charbonneau, D., Brown, T.~M., Latham, D.~W., \& Mayor, M.\ 2000, \apjl, 529, L45 

\bibitem[Childs et al.(2012)]{visitref} Childs, H., Brugger, E., Whitlock, B., et al.\ 2012, in High Performance Visualization: Enabling Extreme-Scale Scientific Insight, ed. E.~W. Bethel et al. (Boca Raton: CRC Press), 357

\bibitem[Cohen et al.(2011)]{cohen} Cohen, O., Kashyap, V.~L., Drake, J.~J., et al.\ 2011, \apj, 733, 67 

\bibitem[Ehrenreich et al.(2008)]{ehrenreich} Ehrenreich, D., Lecavelier des Etangs, A., H{\'e}brard, G., et al.\ 2008, \aap, 483, 933 

\bibitem[Ehrenreich et al.(2015)]{ehrenreich15} Ehrenreich, D., Bourrier, V., Wheatley, P.~J., et al.\ 2015, \nat, 522, 459 

\bibitem[Garc{\'{\i}}a Mu{\~n}oz(2007)]{garciamunoz} Garc{\'{\i}}a Mu{\~n}oz, A.\ 2007, \planss, 55, 1426 

\bibitem[Gendelev \& Krumholz(2012)]{gendelev} Gendelev, L., \& Krumholz, M.~R.\ 2012, \apj, 745, 158 

\bibitem[Gross(1972)]{gross} Gross, S.~H.\ 1972, Journal of Atmospheric Sciences, 29, 214 

\bibitem[Haswell et al.(2012)]{haswell} Haswell, C.~A., Fossati, L., Ayres, T., et al.\ 2012, \apj, 760, 79 

\bibitem[Henry et al.(2000)]{henry} Henry, G.~W., Marcy, G.~W., Butler, R.~P., \& Vogt, S.~S.\ 2000, \apjl, 529, L41 

%
\bibitem[Hubbard et al.(2007)]{hubbard} Hubbard, W.~B., Hattori, M.~F., Burrows, A., Hubeny, I., \& Sudarsky, D.\ 2007, Icarus, 187, 358 

\bibitem[Koskinen et al.(2013)]{koskinen13} Koskinen, T.~T., Harris, M.~J., Yelle, R.~V., \& Lavvas, P.\ 2013, Icarus, 226, 1678 

\bibitem[Koskinen et al.(2014)]{koskinen} Koskinen, T.~T., Lavvas, P., Harris, M.~J., \& Yelle, R.~V.\ 2014, Royal Society of London Philosophical Transactions Series A, 372, 30089 

\bibitem[Krumholz et al.(2007)]{krumholz07f} Krumholz, M.~R., Stone, J.~M., \& Gardiner, T.~A.\ 2007, \apj, 671, 518 

\bibitem[Kulow et al.(2014)]{kulow} Kulow, J.~R., France, K., Linsky, J., \& Loyd, R.~O.~P.\ 2014, \apj, 786, 132 

\bibitem[Lammer et al.(2003)]{lammer} Lammer, H., Selsis, F., Ribas, I., et al.\ 2003, \apjl, 598, L121 

\bibitem[Lecavelier des Etangs et al.(2010)]{lecavelier10} Lecavelier des Etangs, A., Ehrenreich, D., Vidal-Madjar, A., et al.\ 2010, \aap, 514, A72 

\bibitem[Lecavelier des Etangs et al.(2012)]{lecavelier12} Lecavelier des Etangs, A., Bourrier, V., Wheatley, P.~J., et al.\ 2012, \aap, 543, L4 

\bibitem[Li et al.(2010)]{li10} Li, S.-L., Miller, N., Lin, D.~N.~C., \& Fortney, J.~J.\ 2010, \nat, 463, 1054 

\bibitem[Linsky et al.(2010)]{linsky} Linsky, J.~L., Yang, H., France, K., et al.\ 2010, \apj, 717, 1291 

\bibitem[Llama et al.(2013)]{llama} Llama, J., Vidotto, A.~A., Jardine, M., et al.\ 2013, \mnras, 436, 2179 

\bibitem[Llama \& Shkolnik(2015)]{llama15} Llama, J., \& Shkolnik, E.~L.\ 2015, \apj, 802, 41 

\bibitem[Lopez \& Fortney(2013)]{lopez} Lopez, E.~D., \& Fortney, J.~J.\ 2013, \apj, 776, 2 

\bibitem[Matsakos et al.(2015)]{matsakos} Matsakos, T., Uribe, A., {K\"ouml}nigl, A.\ 2015, \aap, 578, A6 

\bibitem[Menager et al.(2013)]{menager} Menager, H., Barth{\'e}lemy, M., Koskinen, T., et al.\ 2013, Icarus, 226, 1709 

\bibitem[Morton(2003)]{morton} Morton, D.~C.\ 2003, \apjs, 149, 205 

\bibitem[Murray-Clay et al.(2009)]{murray} Murray-Clay, R.~A., Chiang, E.~I., \& Murray, N.\ 2009, \apj, 693, 23 

\bibitem[Osterbrock(1989)]{osterbrock} Osterbrock, D.~E.\ 1989, Astrophysics of Gaseous Nebulae and Active Galactic Nuclei (Mill Valley: University Science Books)

\bibitem[Owen \& Jackson(2012)]{owenjackson} Owen, J.~E., \& Jackson, A.~P.\ 2012, \mnras, 425, 2931 


\bibitem[Owen \& Adams(2014)]{owenadams} Owen, J.~E., \& Adams, F.~C.\ 2014, \mnras, 444, 3761 

\bibitem[Parker(1958)]{parker} Parker, E.~N.\ 1958, \apj, 128, 664 

\bibitem[Poppenhaeger et al.(2013)]{poppenhager} Poppenhaeger, K., 
Schmitt, J.~H.~M.~M., \& Wolk, S.~J.\ 2013, \apj, 773, 62 

\bibitem[Quirk(1994)]{quirk} Quirk, J.~J.\ 1994, Int. J. Numer. Methods Fluids, 18, 555 

\bibitem[Rappaport et al.(2014)]{rappaport} Rappaport, S., 
Barclay, T., DeVore, J., et al.\ 2014, \apj, 784, 40 

\bibitem[Schneiter et al.(2007)]{schneiter} Schneiter, E.~M., Vel{\'a}zquez, P.~F., Esquivel, A., Raga, A.~C., \& Blanco-Cano, X.\ 2007, \apjl, 671, L57 

\bibitem[Sofia \& Meyer(2001)]{sofia} Sofia, U.~J., \& Meyer, D.~M.\ 2001, \apjl, 554, L221 

\bibitem[Southworth et al.(2012)]{southworth} Southworth, J., Hinse, T.~C., Dominik, M., et al.\ 2012, \mnras, 426, 1338 

\bibitem[Stone et al.(2008)]{stone} Stone, J.~M., Gardiner, T.~A., Teuben, P., Hawley, J.~F., \& Simon, J.~B.\ 2008, \apjs, 178, 137 

\bibitem[Stone \& Proga(2009)]{proga} Stone, J.~M., \& Proga, D.\ 2009, \apj, 694, 205 

\bibitem[Tenorio-Tagle et al.(1986)]{tenorio} Tenorio-Tagle, G., Bodenheimer, P., Lin, D.~N.~C., \& Noriega-Crespo, A.\ 1986, \mnras, 221, 635 

\bibitem[Tian et al.(2005)]{tian} Tian, F., Toon, O.~B., Pavlov, A.~A., \& De Sterck, H.\ 2005, \apj, 621, 1049 


\bibitem[Trammell et al.(2011)]{trammell} Trammell, G.~B., Arras, P., \& Li, Z.-Y.\ 2011, \apj, 728, 152 

\bibitem[Trammell et al.(2014)]{trammell14} Trammell, G.~B., Li, Z.-Y., \& Arras, P.\ 2014, \apj, 788, 161

\bibitem[Tremblin \& Chiang(2013)]{tremblin} Tremblin, P., \& Chiang, E.\ 2013, \mnras, 428, 2565 

\bibitem[Vidal-Madjar et al.(2003)]{vidal03} Vidal-Madjar, A., Lecavelier des Etangs, A., D{\'e}sert, J.-M., et al.\ 2003, \nat, 422, 143 

\bibitem[Vidal-Madjar et al.(2004)]{vidal04} Vidal-Madjar, A., D{\'e}sert, J.-M., Lecavelier des Etangs, A., et al.\ 2004, \apjl, 604, L69 

\bibitem[Watson et al.(1981)]{watson} Watson, A.~J., Donahue, T.~M., \& Walker, J.~C.~G.\ 1981, Icarus, 48, 150 

\bibitem[Whalen et al.(2004)]{whalen} Whalen, D., Abel, T., \& Norman, M.~L.\ 2004, \apj, 610, 14 

\bibitem[Woods et al.(1998)]{woods} Woods, T.~N., Rottman, 
G.~J., Bailey, S.~M., Solomon, S.~C., \& Worden, J.~R.\ 1998, \solphys, 177, 133 

\bibitem[Yelle(2004)]{yelle} Yelle, R.~V.\ 2004, Icarus, 170, 167 

\end{thebibliography}
\end{document}